\documentclass[twocolumn]{revtex4-2}
\usepackage[dvipdfmx]{graphicx}
\usepackage{amsmath,amsbsy,amssymb}
\usepackage{bm}
\usepackage{mathrsfs}
\usepackage{ulem}
\usepackage{textgreek}
\usepackage{multirow}

\usepackage{color}

\newcommand{\diff}{\mathrm{d}}
\newcommand{\imag}{\mathrm{Im}\,}
\newcommand{\real}{\mathrm{Re}\,}
\newcommand{\trace}{\mathrm{Tr}\,}
\newcommand{\imu}{\mathrm{i}}
\newcommand{\epn}{\mathrm{e}}
\newcommand{\sgn}{\mathrm{sgn}\,}

\newcommand{\dg}{\dagger}
\newcommand{\la}{\langle}
\newcommand{\ra}{\rangle}
\newcommand{\al}{\alpha}

\newcommand{\ep}{\varepsilon}

\begin{document}

\title{
Odd-frequency pairing inherent in
Bogoliubov Fermi liquid
}

\author{
Tatsuya Miki, Shun-Ta Tamura, Shoma Iimura, and Shintaro Hoshino
}

\affiliation{
Department of Physics, Saitama University, Shimo-Okubo, Saitama 338-8570, Japan
}

\date{\today}

\begin{abstract}
The disorder and interaction effects on Bogoliubov-Fermi surfaces with preserved inversion symmetry are studied for a low-energy effective model coupled to bosonic degrees of freedom.
It is shown that the non-ideal Bogoliubov quasiparticles (bogolons) generically induce the odd-frequency pair amplitude which reflects a Cooper pairing at different time.
The self-energy of bogolons is mainly contributed by the disorder effects in the low frequency limit as in the usual electron liquid.
Depending on the choice of the parameters, there are two kinds of solutions: one is frequency-independent (but with sign function of frequency) and the other is proportional to the inverse of the frequency, which exist in both the normal and anomalous parts of the self-energy.
These characteristic self-energy structures are clearly reflected in the single-particle spectrum.
Since the bogolons are originally composed of electrons, the connection between the two is also sought using the concrete $j=3/2$ fermion model, which reveals that the odd-frequency pairing of bogolons is mainly made of the electrons' odd-frequency pairing.
\end{abstract}

\maketitle

The superconductivity is induced by the interactions among electrons near the Fermi surfaces.
In the resultant ground state, the Cooper pair condensation energy is gained by energy gap formation near the Fermi level.
While the Fermi surface usually disappears in the pairing state, they can remain in some superconducting states \cite{Volovik89, Volovik93, Liu03, Gubankova05, Agterberg17, Brydon18, Yuan18, Sumita19, Menke19, Link20, Link20-2, Autti20}, where the elementary excitations near the Fermi surfaces are composed of Bogoliubov quasiparticles (bogolons) \cite{Bogoliubov58}.
For the time-reversal symmetry broken system with preserved inversion symmetry, such Bogoliubov-Fermi surfaces are stable as they are topologically protected \cite{Agterberg17}.
Whereas the bogolon is not a simple charged particle, it can carry energy.
Hence the thermal properties such as specific heat and thermal conductivity are expected to be similar to the conventional Fermi liquid of electrons \cite{Lapp20,Setty20-2} and are potentially observed in the actual materials \cite{Sato18,Setty20,Shibauchi20}.
However, since the bogolons are quasiparticles in the superconducting state, the physical properties should be different from those of the electrons.
Therefore it is desirable to clarify the difference between the Fermi liquid and the {\it Bogoliubov Fermi liquid}, the latter of which is realized for the non-ideal bogolons generically.
Recently, the effect of interactions are considered for bogolons and the possible instabilities are studied \cite{Oh20,Tamura20,Timm20,Herbut20}.
Here, we show that the impurity and correlation effects on this Bogoliubov Fermi surface generate purely odd-frequency pairing amplitude at low energies, which gives a clear distinction from the normal Fermi liquid state of electrons.

The odd-frequency pairing is proposed as a possible exotic ground state of the fermionic systems where the Cooper pair amplitude has an odd function in relative time or frequency, meaning that the pair formation occurs only at different time \cite{Berezinskii74, Kirkpatrick91,Belitz92,Balatsky92,Emery92,Schrieffer94,Bergeret05,Tanaka12,Linder19}.
The actual realizations as a bulk state have been proposed in correlated electron models \cite{Balatsky93,Vojta99,Fuseya03,Yada08,Hotta09,Shigeta09,Shigeta11,Kusunose11-2,Yanagi12} and Kondo lattices \cite{Emery93,Coleman93,Coleman94,Zachar96,Jarrell97,Anders02,Anders02-2,Hoshino14,Hoshino14-2,Otsuki15,Kusunose16,Tsvelik19}.
However, it has been argued that the spatially uniform odd-frequency pairing cannot be a thermodynamically stable state in a conventional framework \cite{Heid95}.
Then, several possibilities to remedy this problem are proposed such as spatially modulated superconducting states \cite{Heid95,Coleman94,Hoshino14} and non-Hermitian description \cite{Belitz99,Solenov09,Kusunose11}. As for the latter scenario, potential problems have also been pointed out \cite{Fominov15}.
In any case, it has been recognized that the spatially uniform and purely odd-frequency pairing state without conventional (even-frequency) pairing cannot simply be realized.
On the other hand, such exotic pairing has been discussed also in edge or interface \cite{Bergeret01,Eschrig03,Tanaka07-1,Tanaka07-2}, which are secondarily induced from the even-frequency pairing in bulk.

\begin{figure}[b]
\begin{center}
\includegraphics[width=60mm]{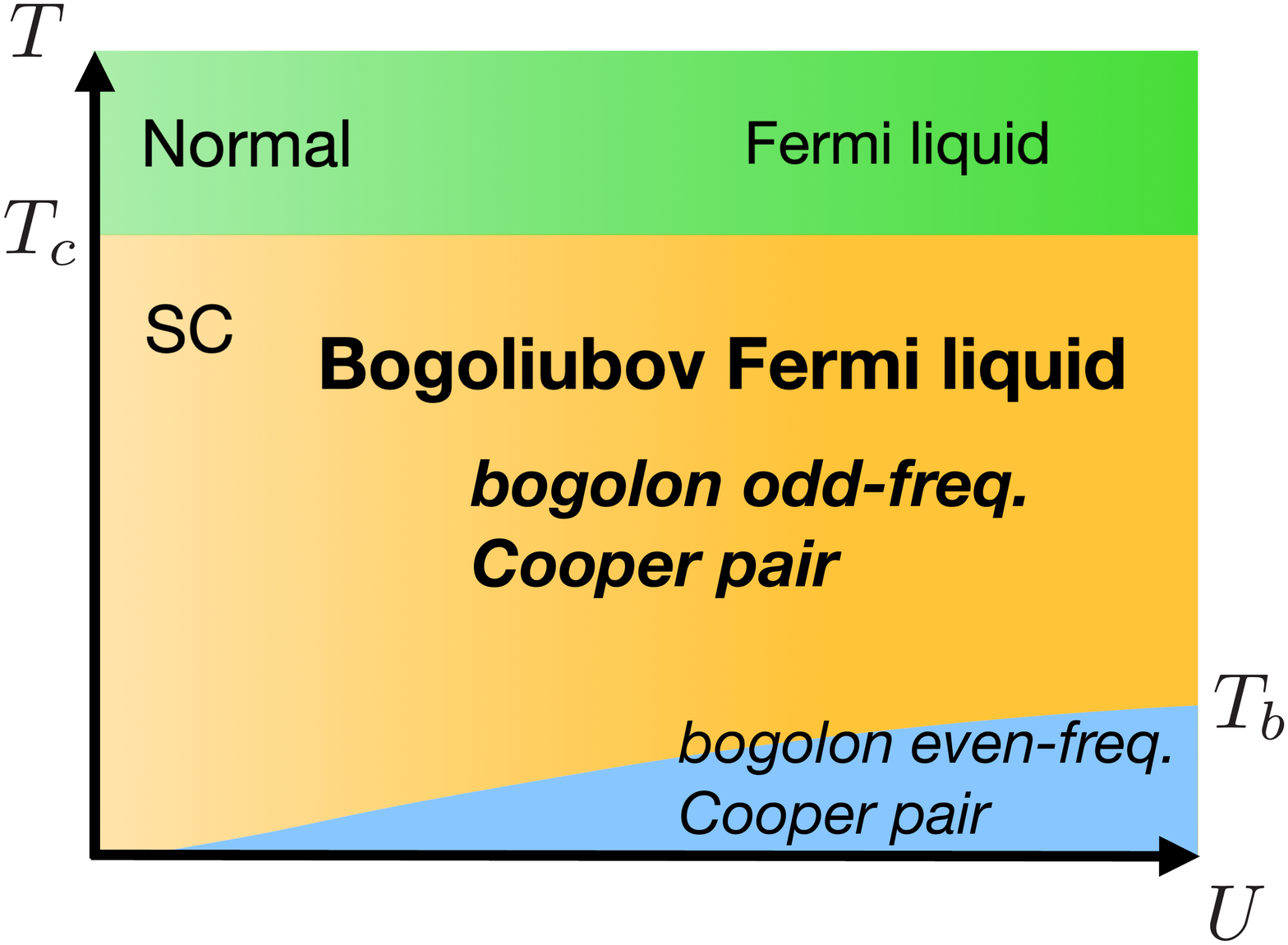}
\caption{
Schematic phase diagram for the Bogoliubov Fermi liquid.
The horizontal axis with $U$ indicates the deviation from the non-interacting bogolons.
}
\label{fig:concept}
\end{center}
\end{figure}

In this paper, we demonstrate the emergence of the spatially uniform odd-frequency pair in the world not of electrons but of bogolons.
This odd-frequency pairing is induced by a non-ideality of bogolons, where the broken gauge-symmetry plays a crucial role as different from the electronic interactions.
As shown in the following, the induced odd-frequency pair is closely related to the energy structure of the usual Fermi-liquid self-energy existing in the diagonal Green function for electrons, but for bogolons it is reflected also in the off-diagonal part, i.e., pair amplitude and pair potential.

In the previous works \cite{Oh20,Tamura20,Timm20,Herbut20}, the possibility of the spontaneous symmetry breaking induced by interactions among bogolons is considered.
Here, we concentrate on the superconducting state above such low-temperature phase ($T > T_b$) but below the superconducting transition temperature ($T<T_c$).
Namely, we deal with a ``normal state'' of bogolons.
Our situation is sketched in Fig.~\ref{fig:concept}.

Below, we connect the two important concepts of superconductivity, i.e., the Bogoliubov Fermi surface and odd-frequency pairing, based on the concrete model.
As in the theories of superconductivity and of Fermi liquid, the degrees of freedom near the Fermi level are important for the low-temperature and low-energy properties.
Hence we also assume that the dominant contribution enters through the degrees of freedom near the Bogoliubov Fermi surface.
The non-interacting part of the Hamiltonian for bogolons  near the Fermi level is written as
\begin{align}
  \mathscr H_0 &= \sum_{\bm k} \xi_{\bm k} \al_{\bm k}^\dg \al_{\bm k}.
  \label{eq:ham_zero}
\end{align}
In the following, we consider the inversion symmetric systems ($\xi_{-\bm k}=\xi_{\bm k}$) with a time-reversal symmetry breaking where the Fermi surface is topologically protected \cite{Agterberg17}.
In order to have an intuition for the energy scales, we write the energy dispersion as $\xi_{\bm k} = \frac{\bm k^2}{2m_b} - \ep_{{\rm F}b}$ for simplicity, where $m_b$ and $\ep_{{\rm F}b}$ are effective mass and Fermi energy for bogolons ($\hbar=k_{\rm B}=1$).
As inferred from the original electronic system \cite{Agterberg17} on which bogolons are based, each quantity is roughly expressed as $m_{b} \sim \frac{\Delta}{\ep_{{\rm F}e}} m_e$ and $\ep_{{\rm F}b} \sim \Delta$ where $m_e$ and $\ep_{{\rm F}e}$ are mass and Fermi energy for electrons, and $\Delta$ is the energy scale for the superconducting gap function. Here we have assumed the magnitude relation $\ep_{{\rm F}e} > \Delta$.
The Fermi wavenumber is given by $k_{{\rm F}b} = \sqrt{2m_b \ep_{{\rm F}b}}$.
Thus the Fermi velocity for bogolons is similar to that of electrons: $v_{{\rm F}b} = \frac{k_{{\rm F}b}}{m_b} \sim v_{{\rm F}e}$.
Note that the potential term proportional to $\al^\dg_{\bm k} \al^\dg_{-\bm k}$ cannot exist together with the inversion symmetry.
Such pair potential may be generated with spontaneous symmetry breaking below the ordering temperature $T_b$ for bogolons as shown in Fig.~\ref{fig:concept}, although we work in the temperature regime $T>T_b$ in the following.

For the superconducting state considered here, the time-reversal and gauge symmetries are already broken.
Nevertheless, in terms of bogolons as in Eq.~\eqref{eq:ham_zero}, the effects from these broken symmetries are not described apparently.
This paradox at first sight is rationalized by considering the self-energy terms in which the symmetry of the system is reasonably reflected.
Then, we arrive at the important conclusion that the anomalous part can be finite in the presence of correlations, and its symmetry must be even-parity pairing.
Since the bogolons have no internal degrees of freedom at the Fermi level, and in order to satisfy the Pauli principle, the pairing state of bogolons must have odd relative time dependence.
Hence the odd-frequency pairing is reasonably realized.

We proceed along with the conventional microscopic approach for the Fermi liquid based on the weak-coupling limit \cite{AGD_book}.
The total Hamiltonian is written as $\mathscr H =\mathscr H_0 + \mathscr H_{\rm imp} + \mathscr H_{\rm int}$, where the impurity scattering part is given by
\begin{align}
  \mathscr H_{\rm imp} &= \frac 1 V\sum_{\bm k , \bm q} \Big[
  \rho_{\bm q} u_1(\bm k,\bm q)
  \al^\dg_{\bm k+\bm q} \al_{\bm k}
  \nonumber \\
  &\hspace{15mm}
  + \rho_{\bm q} u_2(\bm k,\bm q)
  \al^\dg_{\bm k+\bm q} \al^\dg_{-\bm k}
  +{\rm H.c.}
  \Big],
  \label{eq:imp}
\end{align}
where $\rho_{\bm q} = \sum_{i} \epn^{\imu \bm q\cdot \bm{R}_i}$ is the structure factor for the impurity configuration $\{ \bm{R}_i \}$, and $V$ is a system volume.
The second term with $\al^\dg\al^\dg$ is characteristic of the bogolon systems.
We also consider the correlation effects in $\mathscr H_{\rm int}$.
Among the various interactions, we take the simple model where the bogolons are coupled with bosons.
The interaction term is given by the replacement
\begin{align}
  u_j(\bm k,\bm q) \rho_{\bm q}
  \to  \imu
  g_j(\bm k,\bm q) \sqrt{\frac{\omega_{0,\bm q}}{2}} (b_{\bm q} - b_{-\bm q}^\dg) \label{eq:int_boson}
\end{align}
in Eq.~\eqref{eq:imp}, where $g_{j=1,2}$ is the coupling constant, $\omega_{0,\bm q}$ the bare boson dispersion, and $b$ ($b^\dg$) the annihilation (creation) operator of boson.
Note that, in the above model, we have assumed that the Bogoliubov Fermi surfaces are stable as separated from the other bands located at higher energies.
Then, the disorders and interactions dominantly affect the bogolons located near the Bogoliubov Fermi surfaces.
Otherwise, the original assumption of the stable Bogoliubov Fermi surfaces must be reconsidered.

\begin{figure}[t]
\begin{center}
\includegraphics[width=75mm]{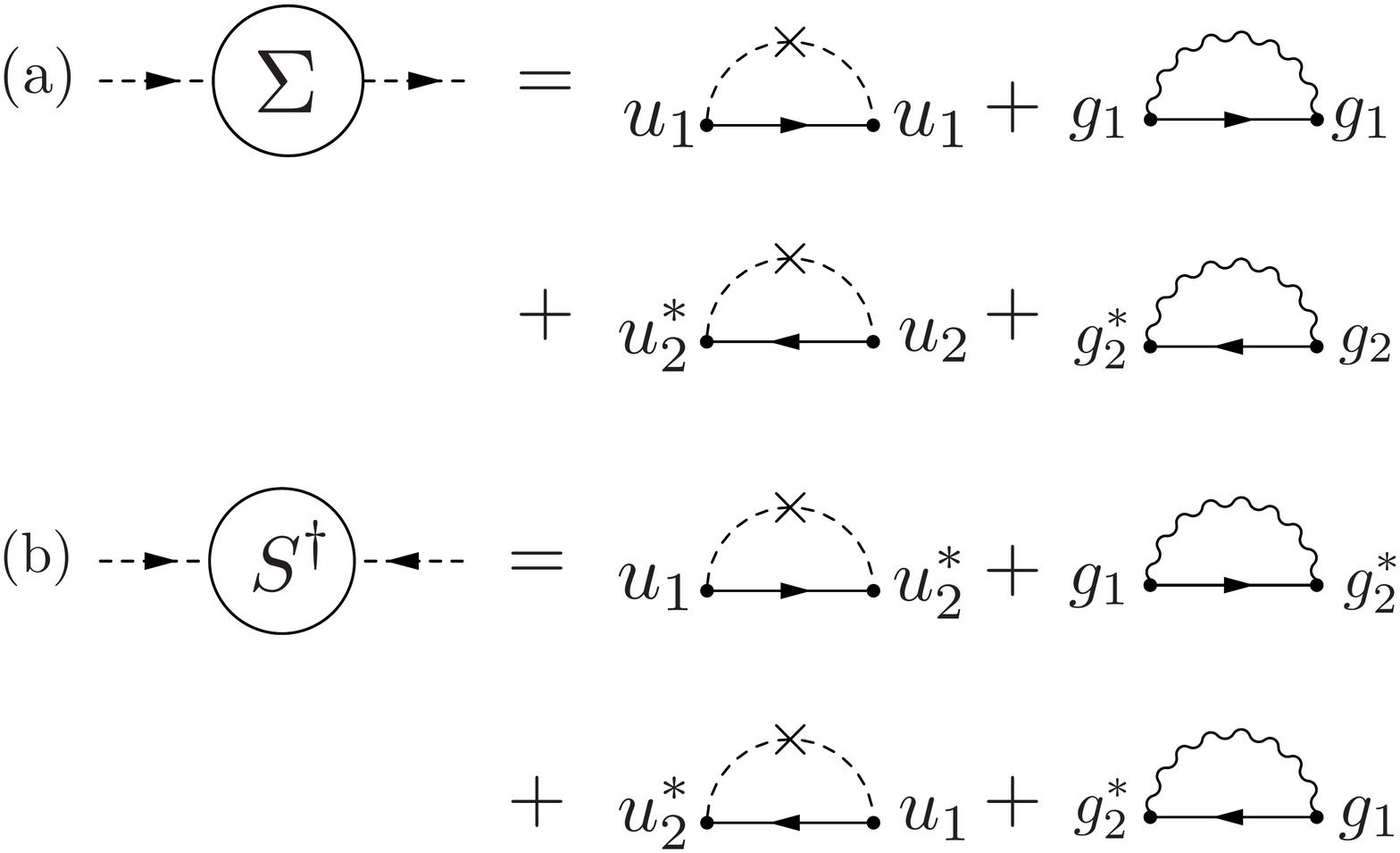}
\caption{
Diagrams for (a) normal and (b) anomalous self-energies which are relevant to the Bogoliubov Fermi liquid.
}
\label{fig:diagram}
\end{center}
\end{figure}

In order to show the presence of the pure odd-frequency pairing,
we use the simple weak-coupling perturbation theory, which is a minimal description for the Fermi liquid and is sufficient for generating  anomalous Green functions.
The self-energy is contributed by the diagrams shown in Fig.~\ref{fig:diagram}, and their calculation is parallel to the usual Fermi liquid theory \cite{AGD_book} and is shown in Supplementary Material (SM) A and B \cite{SM}.
The resultant normal part of the self-energy is given by
\begin{align}
  \Sigma_{\bm k}(\imu\ep_n) &= - \imu\Gamma_{1\bm k} {\rm sgn\,}\ep_n +  a_{\bm k} \imu \ep_n
  + \imu b_{\bm k}(\pi^2T^2 - \ep_n^2) {\rm sgn\,}\ep_n,
  \label{eq:se_diag}
\end{align}
where $\ep_n = (2n+1)\pi T$ is the fermionic Matsubara frequency.
The first term corresponds to a quasiparticle damping due to impurity scattering.
The second term represents a renormalization factor, and the third term is responsible for the damping in the usual Fermi liquid theory.
For bogolons, the anomalous self-energy is also present, whose diagrammatic contribution is very similar to the normal self-energy as shown in Fig.~\ref{fig:diagram}.
The anomalous part is obtained as
\begin{align}
  S^\dg
  _{\bm k}(\imu\ep_n) &= - \imu\Gamma_{2\bm k} {\rm sgn\,}\ep_n +  c_{\bm k} \imu \ep_n
  + \imu d_{\bm k}(\pi^2T^2 - \ep_n^2) {\rm sgn\,}\ep_n,
  \label{eq:se_offd}
\end{align}
which is the same frequency dependence as the usual Fermi liquid.
Namely, the Fermi-liquid self-energy is originally odd in frequency and therefore it matches well with the requirement of the bogolon anomalous self-energy under the inversion symmetry.
This is the reason why the odd-frequency pair potential naturally appears for bogolons.
We emphasize that the {\it spatially uniform and purely odd-frequency pairing} is realized in the present setup, which is necessarily accompanied by the normal self-energies.

From the Hermiticity relation, it can be shown that $a$, $b$ and $\Gamma_1$ are real, while $c$,  $d$ and $\Gamma_2$ can be complex.
As discussed in Ref.~\cite{Agterberg17}, the Bogoliubov Fermi surfaces can be realized for a chiral $d$-wave superconductivity with the gap function $\Delta_{\bm k}\sim k_z (k_x + \imu k_y)$ of the original electrons.
In this case, the anomalous part $c$ (and also $d$) may include the contribution $\propto k_z (k_x + \imu k_y)$ from the symmetry argument.
Hence the broken gauge symmetry and time-reversal symmetry are clearly reflected in the anomalous self-energy of bogolons which is odd in frequency.

The effect of the presence of the odd-frequency pairing is best visualized in the single-particle spectral functions.
To see this, we consider the Green functions given by
\begin{align}
  \begin{pmatrix}
  G_{\bm k} & F_{\bm k}\\
  F^\dg_{\bm k} & \bar G_{\bm k}
  \end{pmatrix}
  ^{-1}
  &=
  \begin{pmatrix}
  \imu\ep_n -  \xi_{\bm k} & \\
   & \imu \ep_n + \xi_{\bm k}
  \end{pmatrix}
  -
  \begin{pmatrix}
  \Sigma_{\bm k} & S_{\bm k} \\
  S^\dg_{\bm k} & \bar \Sigma_{\bm k}
  \end{pmatrix},
\end{align}
where the frequency dependence is omitted.
The self-energies satisfy the relations
$\bar \Sigma_{\bm k}(\imu\ep_n) = - \Sigma_{\bm k}^*(\imu\ep_n)$ and $S^\dg_{\bm k}(\imu\ep_n) = - S_{\bm k}^*(\imu \ep_n)$ as derived from the Hermiticity and inversion symmetry.
The dominant contribution of the self-energies at low energies enters from the impurity effect.
Hence here we focus on the impurity self energies $\Gamma_{1,2}$ in Eqs.~\eqref{eq:se_diag} and \eqref{eq:se_offd}.
We note that $\Gamma_1 > |\Gamma_2|$ is required for the physical behavior, i.e., the positive weight of the spectrum, and it can indeed be checked in the weak coupling limit \cite{SM}.
In the following, we neglect the $\bm k$-dependence in the self-energy for simplicity, corresponding to the spatially local self-energies.

\begin{figure}[t]
\begin{center}
\includegraphics[width=85mm]{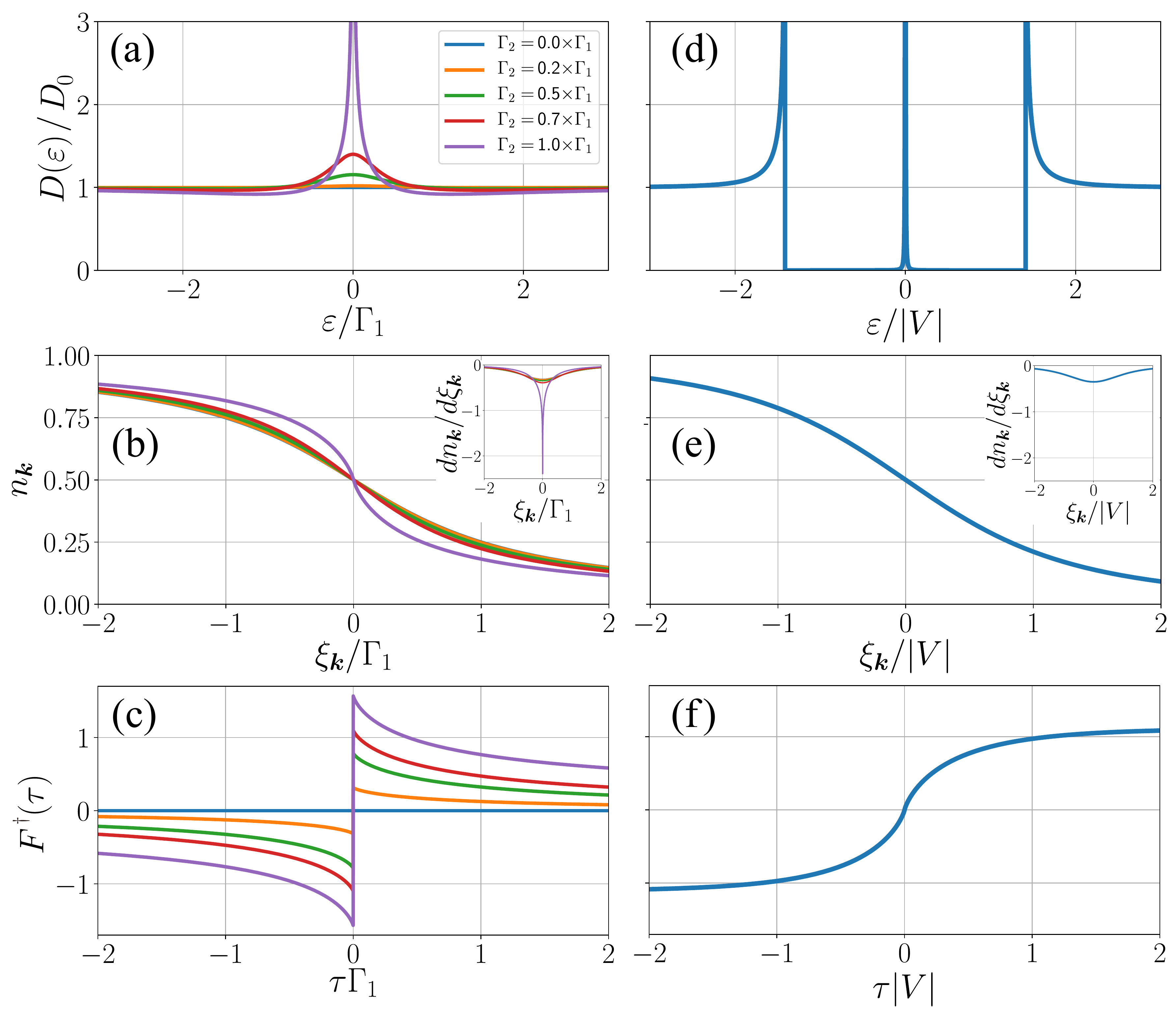}
\caption{
(a) Density of states, (b) momentum distribution function, (c) pair potential for the solution of the first kind.
(d,e,f) are similar to (a,b,c), respectively, but for the solution of the second kind.
$D_0$ is a bare density of states.
The pair amplitude in (c) is normalized by $\Gamma_1$ and in (d) by $|V|$ where the phases are chosen as zero.
}
\label{fig:spectrum}
\end{center}
\end{figure}

The single particle spectrum $A_{\bm k}(\ep)$ is obtained from the imaginary part of the retarded Green function.
The density of states is then evaluated by performing the $\bm k$-integral as $D(\ep) = \sum_{\bm k} A_{\bm k}(\ep)$, which is shown in Fig.~\ref{fig:spectrum}(a).
The larger $\Gamma_2$ makes the higher peak at the Fermi level.
Clearly, the presence of odd-frequency pair is responsible for this characteristic feature near the Fermi level.
Figure~\ref{fig:spectrum}(b) shows the momentum distribution function $n_{\bm k} = \la \al_{\bm k}^\dg \al_{\bm k} \ra$.
The sharp drop at the Fermi energy in the ideal limit is smeared by the damping $\Gamma_1$, but is recovered with increasing $|\Gamma_2|$.
The inset of (b) shows the derivative of this function, where the change becomes more abrupt for the larger pair potential $\Gamma_2$ and diverges when $|\Gamma_2| \to \Gamma_1$.

We also show the spatially local anomalous pair amplitude $F = \sum_{\bm k} F_{\bm k}$.
Performing the Fourier transformation, we show in Fig.~\ref{fig:spectrum}(c) the imaginary-time $\tau$ dependence of the pair amplitude where we have taken the zero temperature limit.
The functional form clearly shows the odd-frequency pairing.
The value of the pair amplitude is largest at short time and is discontinuous at $\tau=0$.
The asymptotic behavior at long time is $F^\dg(\tau \to \infty) \sim \tau^{-1}$.

The information of quasiparticles are seen in the real-time evolution of the retarded Green function $G^{\rm R}_{\bm k}(\ep) = G_{\bm k}(\ep+\imu 0^+)$ \cite{SM}.
We can explicitly evaluate it in the real-time domain and obtain $G_{\bm k}^{\rm R} (t) \sim \epn^{\imu z_{\bm k} t}$.
For the high-energy region with $\xi_{\bm k}>|\Gamma_2|$, the complex energy is given by $z_{\bm k} = \pm \sqrt{\xi_{\bm k}^2 - |\Gamma_2|^2} + \imu \Gamma_1$.
This is a standard form composed of the oscillating part with the quasiparticle energy $\sqrt{\xi_{\bm k}^2 - |\Gamma_2|^2}$ and the damping $\Gamma_1$.
On the contrary, for $\xi_{\bm k} < |\Gamma_2|$, we have $z_{\bm k} = \imu (\pm \sqrt{|\Gamma_2|^2- \xi_{\bm k}^2} + \Gamma_1)$ which is pure imaginary.
Hence the low-energy part has no oscillating part and has only damping with two relaxation rates.

For the impurity effect, we have also analyzed the equations with the self-consistent treatment \cite{SM}.
We have found that the frequency-independent $\Gamma_{1,2}$ solution discussed above is not much modified at low energies.
This type is called the first-kind.
On the other hand, owing to the non-linearity of the equations, we have also found another type of solution depending on the choice of the parameters.
This solution of the second-kind has the frequency dependence
\begin{align}
  \Sigma(\imu\ep_n) = \frac{|V|^2}{\imu\ep_n}
  ,\ \ \
  S^\dg (\imu\ep_n) = \frac{V^{2}}{\imu\ep_n}
  \label{eq:second_kind}
\end{align}
at low energies, where $V$ is a complex constant.
In this case, we have the three single-particle excitation energies $E_{\bm k} = 0, \pm \sqrt{\xi_{\bm k}^2 + 2|V|^2}$ from the pole of Green functions.
This energy structure is due to the fact that the same absolute value of $|V|$ is shared for both normal and anomalous part of self-energy.
The presence of the gap structure and the zero-energy peak at the Fermi level is the characteristic feature for the density of states as shown in Fig.~\ref{fig:spectrum}(d), where the weight of the zero-energy peak is proportional to $|V|$.
Figure~\ref{fig:spectrum}(e,f) show the momentum distribution function and imaginary-time dependence of local pair amplitude, respectively, for the case with Eq.~\eqref{eq:second_kind}.
The pair amplitude has odd function form but is now smooth at equal imaginary-time.
The asymptotic behaviors are identified as
$F^\dg(\tau) \sim -|V|^2\tau\ln(\tau|V|) $
for $\tau\to 0$ and $F^\dg(\tau) \sim |V|$
for $\tau \to \infty$ \cite{SM}.

The frequency dependence proportional to the inverse of $\ep_n$ for both the normal and anomalous parts are the same features as the pairing states in the multichannel Kondo lattices where the conduction electrons hybridized virtually with the localized fermions \cite{Coleman94,Coleman99, Flint08, Hoshino14-2, Iimura19}.
This suggests that the $1/\ep_n$ form is a ubiquitous form of the self-energy for odd-frequency pairing in bulk, as it is found in the very different two physical systems.

Thus, the Bogoliubov Fermi liquid shows characteristic single-particle excitations at low energies, which can be much different from the usual electron liquids.
The above two different behaviors, the first- and second-kind solutions, are observed depending on the choice of parameters \cite{SM}, and hence are dependent on the specific materials.
For example, in Fe(Se,S), the Fermi-liquid-like behavior is observed below the superconducting transition temperature as probed by thermal measurement \cite{Sato18,Shibauchi20}.
The solution of the second kind shows the presence of the localized level at the Fermi level, which should not contribute to the transport phenomena.
Hence, the solution of the first-kind with the form \eqref{eq:se_offd} is likely realized in Fe(Se,S).
While the observed behaviors are similar to the Fermi liquid, which implies weak disorder effects, the tuning of the system may make it clearer to detect the fingerprints of Bogoliubov-Fermi liquids.

\begin{figure}[t]
\begin{center}
\includegraphics[width=85mm]{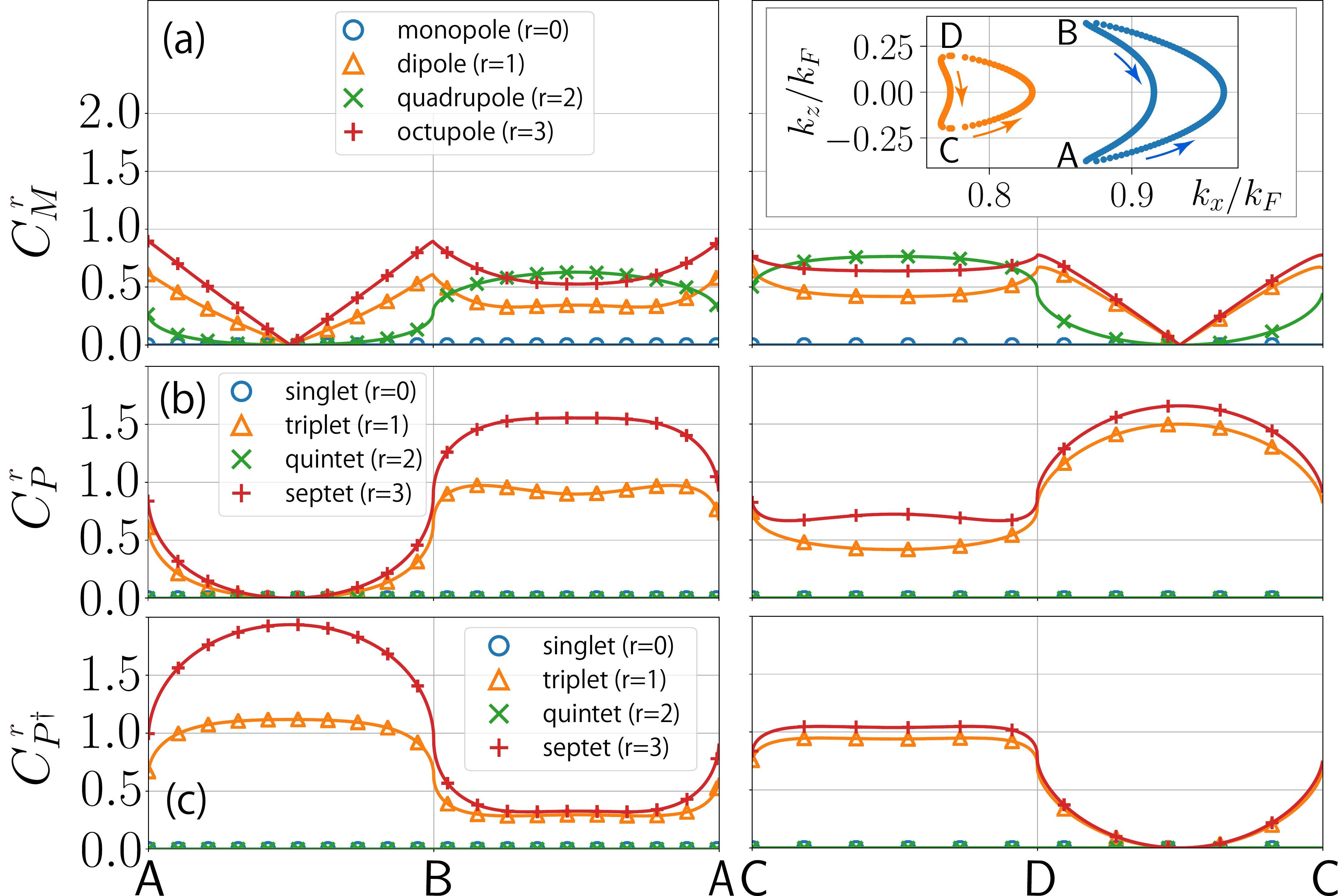}
\caption{
Wavevector dependence of (a) $C_{M}(\bm k)$ (multipoles), (b) $C_{P}(\bm k)$, and (c) $C_{P^\dg}(\bm k)$ (multiplet electron and hole pairs) along the Fermi surface in the $j=3/2$ fermion model.
The inset of (a) shows the two Fermi surfaces near the $k_x$ axis located in the $k_y=0$ plane.
The horizontal axis of (a--c) is a path shown by arrows in the inset of (a).
}
\label{fig:component}
\end{center}
\end{figure}

Finally we discuss the connection of the bogolon degrees of freedom to original electrons.
As a simple realization of the Bogoliubov Fermi surface, we take $j=3/2$ fermion model proposed in Ref.~\cite{Agterberg17}, where the symmetric spin-orbit interaction and time-reversal-symmetry-broken $d+\imu d$ pairing are considered.
The pair amplitude of bogolons is connected to the order parameters of the original electrons.
At each $\bm k$ point on the Fermi surface, we can define the electronic multipoles $M^r_{\bm k}$ and multiplet pair amplitudes $P^r_{\bm k}, P_{\bm k}^{\dg, r}$ (see SM C \cite{SM} for the concrete forms), which are classified by the rank $r$ defined up to $2j$.
For $M$ ($P,P^\dg$), the rank $r=0,1,2,3$ respectively correspond to monopole (singlet pair), dipole (triplet pair), quadrupole (quintet pair), octupole (septet pair) \cite{Tamura20}.
The order parameters induced by the odd-frequency pair amplitude of bogolons are given by
\begin{align}
  X^r_{\bm k} (\tau) &= C^r_X(\bm k) F_{\bm k} (\tau)
\end{align}
for $X = M,P,P^\dg$.
The quantity $C$ is regarded as a kind of `susceptibility', showing how much of multipoles and multiplet pair are induced from the odd-frequency pair amplitude.
For the multiplet pairs, only the spin-triplet and spin septet pair can be finite due to the odd-function in time in systems with inversion symmetry.

Figure \ref{fig:component} shows the value of $C_X$ along the Fermi surface.
For exemplary demonstration, we take the parameters as the symmetric spin-orbit coupling $\beta/\ep_{\rm Fe} = 0.3$, $s$- and $d$-wave pair potentials $\Delta_0/\ep_{\rm Fe} = \Delta_1/\ep_{\rm Fe} = 0.1$ in Refs.~\cite{Agterberg17,Tamura20} (see also SM C \cite{SM}).
The inset of Fig.~\ref{fig:component}(a) shows the Fermi surfaces in the $k_y=0$ plane.
We have two Fermi surfaces near the $xy$-plane, and it has a donut-like shape because of the rotational symmetry around $z$-axis.
The figures (a--c) show that the dominantly induced components are odd-frequency (spin-triplet and spin-septet) electron/hole pair amplitude ($P,P^\dg$).
Namely, the bogolon odd-frequency pair is mainly composed of the electron odd-frequency pair.
We note that the diagonal multipole components [$M$ in (a)] also mix with a similar-order of magnitudes.

To summarize, we have demonstrated that the non-ideal bogolons generate purely odd-frequency pair potential and pair amplitude near the Fermi surface, as different from the usual Fermi liquid of electrons.
The odd-frequency pair recovers the original symmetry of the electron system in terms of bogolons, where the broken gauge and time-reversal symmetries are not reflected in the level of the non-interacting Hamiltonian of bogolon.
The effect of the odd-frequency pairing is clearly seen in the single-particle spectral functions at low energies where the disorder effect is dominantly present.
The system with Bogoliubov Fermi surfaces is a suitable playground for studying the properties of odd-frequency pairing.

\section*{Acknowledgment}
This work was supported by JSPS KAKENHI Grants
No.~JP18K13490,
No.~JP18H01176,
No.~JP18H04305 and
No.~JP19H01842.

\clearpage
\appendix

\makeatletter
\renewcommand{\thepage}{S\arabic{page}}
\renewcommand{\theequation}{S\arabic{equation}}
\renewcommand{\thefigure}{S\arabic{figure}}
\renewcommand{\thetable}{S\arabic{table}}
\makeatother

\setcounter{page}{1}
\setcounter{equation}{0}
\setcounter{table}{0}
\setcounter{figure}{0}

\noindent
{\bf SUPPLEMENTARY MATERIAL FOR \\
``Odd-frequency pairing inherent in
Bogoliubov Fermi liquid''}
\\[2mm]
T. Miki, S.-T. Tamura, S. Iimura, and S. Hoshino
\\[2mm]
(Dated: \today)

\section*{SM A. Impurity effects}

We consider the impurity potential with the form
\begin{align}
  \mathscr H_{\rm imp} &=
  \frac{1}{V}\sum_{\bm k , \bm q}
  \rho_{\bm q}
  u_1(\bm k,\bm q) \al^\dg_{\bm k+\bm q} \al_{\bm k}
  \nonumber \\
  &+ \frac{1}{V} \sum_{\bm k , \bm q}
  \rho_{\bm q}
  u_2 (\bm k, \bm q) \al^\dg_{\bm k+\bm q} \al^\dg_{-\bm k}  + {\rm H.c.},
\end{align}
which breaks the translational symmetry or momentum conservation.
The information for the disorder is included in the structure factor
\begin{align}
  \rho_{\bm q} = \sum_i \epn^{-\imu \bm q\cdot \bm R_i},
\end{align}
where the summation with respect to $i$ is performed for the impurity positions $\bm R_i$, and $n_{\mathrm{imp}} = V^{-1}\sum_{i} 1$.
This quantity satisfies
\begin{align}
  \overline{\rho_{\bm q}\rho_{\bm q'}} = Vn_{\mathrm{imp}}\delta_{\bm q, -\bm q'},
\end{align}
where the overline indicates the average over the impurity configurations.
The connection to the original electrons is shown in SM C.
We have the relations
\begin{align}
  & u_1(\bm k,\bm q) = u_1^* (\bm k+\bm q, -\bm q), \label{eq:hermite_1}
  \\
  & u_2(\bm k,\bm q) = - u_2(-\bm k-\bm q, \bm q), \label{eq:hermite_2}
  \\
  & u_{1,2}(\bm k,\bm q) = u_{1,2} (-\bm k,-\bm q), \label{eq:inversion}
\end{align}
each of which originates from the Hermiticity, Pauli principle (anticommutation relation), and the inversion symmetry, respectively.

The single-particle Green function is given by the $2\times 2$ matrix as
\begin{align}
  \hat G_{\bm k}(\tau) &= - \la \mathcal T
  \begin{pmatrix}
  \al_{\bm k}(\tau) \\
  \al^\dg_{-\bm k}(\tau)
  \end{pmatrix}
  \begin{pmatrix}
  \al_{\bm k}^\dg & \al_{-\bm k}
  \end{pmatrix}
  \ra,
\end{align}
where $\mathcal T$ indicates a imaginary time ordering, and $A(\tau) = \epn^{\tau\mathscr H} A \epn^{-\tau\mathscr H}$ is the Heisenberg picture.
The Fourier transformation is defined by
\begin{align}
  \hat G_{\bm k}(\imu\ep_n) &= \int_0^{1/T} \diff \tau \hat G_{\bm k} (\tau) \epn^{\imu\ep_n \tau}.
\end{align}
As given in the main text, the normal and anomalous self-energies are introduced by
\begin{align}
  \hat G_{\bm k}^{-1}(\imu\varepsilon_n) &=
  \begin{pmatrix}
  \imu\ep_n - \xi_{\bm k} -\Sigma_{\bm k}(\imu\varepsilon_n)  &  -S_{\bm k}(\imu\varepsilon_n)  \\
  -S^\dg_{\bm k}(\imu\varepsilon_n) &  \imu\ep_n + \xi_{\bm k} -\bar \Sigma_{\bm k}(\imu\varepsilon_n)
  \end{pmatrix}. \label{eq:G}
\end{align}
In the following of this section, we will derive explicit form of self-energies.

\subsection*{1. Born approximation}

We consider the lowest-order contribution of the impurity potential.
Second order self-energies (Born approximation) are given as follows:
\begin{align}
  \Sigma_{\bm{k}}(\imu\varepsilon_n) &= \Sigma_{1,\bm{k}}(\imu\varepsilon_n) + 2\Sigma_{2,\bm{k}}(\imu\varepsilon_n), \\
  \Sigma_{1,\bm{k}}(\imu\varepsilon_n) &= n_{\mathrm{imp}} \int \frac{\diff\bm{q}}{(2\pi)^3} |u_1(\bm{q} , \bm{k} - \bm{q})|^2 G^0_{\bm{q}}(\imu\varepsilon_n), \\
  \Sigma_{2,\bm{k}}(\imu\varepsilon_n) &= -2n_{\mathrm{imp}} \int \frac{\diff\bm{q}}{(2\pi)^3} |u_2(\bm{q} , \bm{k} - \bm{q})|^2 G^0_{\bm{q}}(-\imu\varepsilon_n), \\
  S^{\dagger}_{\bm{k}}(\imu\varepsilon_n) &= -2n_{\mathrm{imp}} \int \frac{\diff\bm{q}}{(2\pi)^3} u_1(\bm{q} , \bm{k} - \bm{q}) u_2^\ast(\bm{q} , \bm{k} - \bm{q}) \notag \\
  &\quad \times[G^0_{\bm{q}}(-\imu\varepsilon_n) - G^0_{\bm{q}}(\imu\varepsilon_n)],
\end{align}
which corresponds to the diagrams shown in Fig.~\ref{fig:diagram} of the main text.
For the evaluation of the wavevector integral,
we replace the coupling constant by its averaged value with respect to $\bm q$.
Then we only have to evaluate the integration as follows:
\begin{align}
  \int \frac{\diff\bm{q}}{(2\pi)^3} G^0_{\bm{q}}(\pm \imu\varepsilon_n) &= \int \frac{\diff\bm{q}}{(2\pi)^3} \frac{1}{\pm \imu \varepsilon_n - \xi_{\bm{q}}} \notag \\
  &= \mp \imu\pi D_0 \sgn \varepsilon_n,
\end{align}
where $D_0$ is a density of states of bogolon at Fermi level.
Then,
\begin{align}
  \Sigma_{1,\bm{k}}(\imu\varepsilon_n) &= -\imu\pi n_{\mathrm{imp}} D_0 \sgn \varepsilon_n
  \langle |u_1(\bm{q} , \bm{k} - \bm{q})|^2 \rangle_{\bm q}, \label{eq:sigma_born_1} \\
  \Sigma_{2,\bm{k}}(\imu\varepsilon_n) &= -2\imu\pi n_{\mathrm{imp}} D_0 \sgn \varepsilon_n \langle |u_2(\bm{q} , \bm{k} - \bm{q})|^2 \rangle_{\bm{q}}, \label{eq:sigma_born_2} \\
  S_{\bm{k}}^\dagger(\imu\varepsilon_n) &= -4\imu\pi n_{\mathrm{imp}} D_0 \sgn \varepsilon_n \langle u_1(\bm{q} , \bm{k} - \bm{q}) u_2^\ast(\bm{q} , \bm{k} - \bm{q}) \rangle_{\bm{q}}, \label{eq:s_born}
\end{align}
where $\la \cdots \ra_{\bm q} = \int \diff \bm q \cdots / \int \diff \bm q 1$.
Hence, we obtain the self-energies
\begin{align}
  \Sigma_{\bm{k}}(\imu\varepsilon_n) &= -\imu\Gamma_{1,\bm k} \sgn \varepsilon_n, \\
  S_{\bm{k}}^\dagger(\imu\varepsilon_n) &= -\imu \Gamma_{2,\bm k} \sgn \varepsilon_n,
\end{align}
where
\begin{align}
  \Gamma_{1,\bm{k}} &= \pi n_{\mathrm{imp}} D_0 \big(\langle |u_1(\bm{q} , \bm{k} - \bm{q})|^2 \rangle_{\bm q} + 4\langle |u_2(\bm{q} , \bm{k} - \bm{q})|^2 \rangle_{\bm{q}} \big), \\
  \Gamma_{2,\bm{k}} &= 4\pi n_{\mathrm{imp}} D_0 \langle u_1(\bm{q} , \bm{k} - \bm{q}) u_2^\ast(\bm{q} , \bm{k} - \bm{q}) \rangle_{\bm{q}}.
\end{align}
One sees that the usual Born approximation result is reproduced if $u_2$, which is specific to bogolons, is set to zero.

As pointed out in the main text, the spectral function shows unphysical behavior for $|\Gamma_{2,\bm k}| > \Gamma_{1,\bm k}$.
However, we can show that the condition $|\Gamma_{2,\bm k}| < \Gamma_{1,\bm k}$ is satisfied based on the expressions obtained above together with the magnitude relation between arithmetic and geometric means.

\subsection*{2. Self-consistent treatment}

\begin{figure}[t]
\begin{center}
\includegraphics[width=75mm]{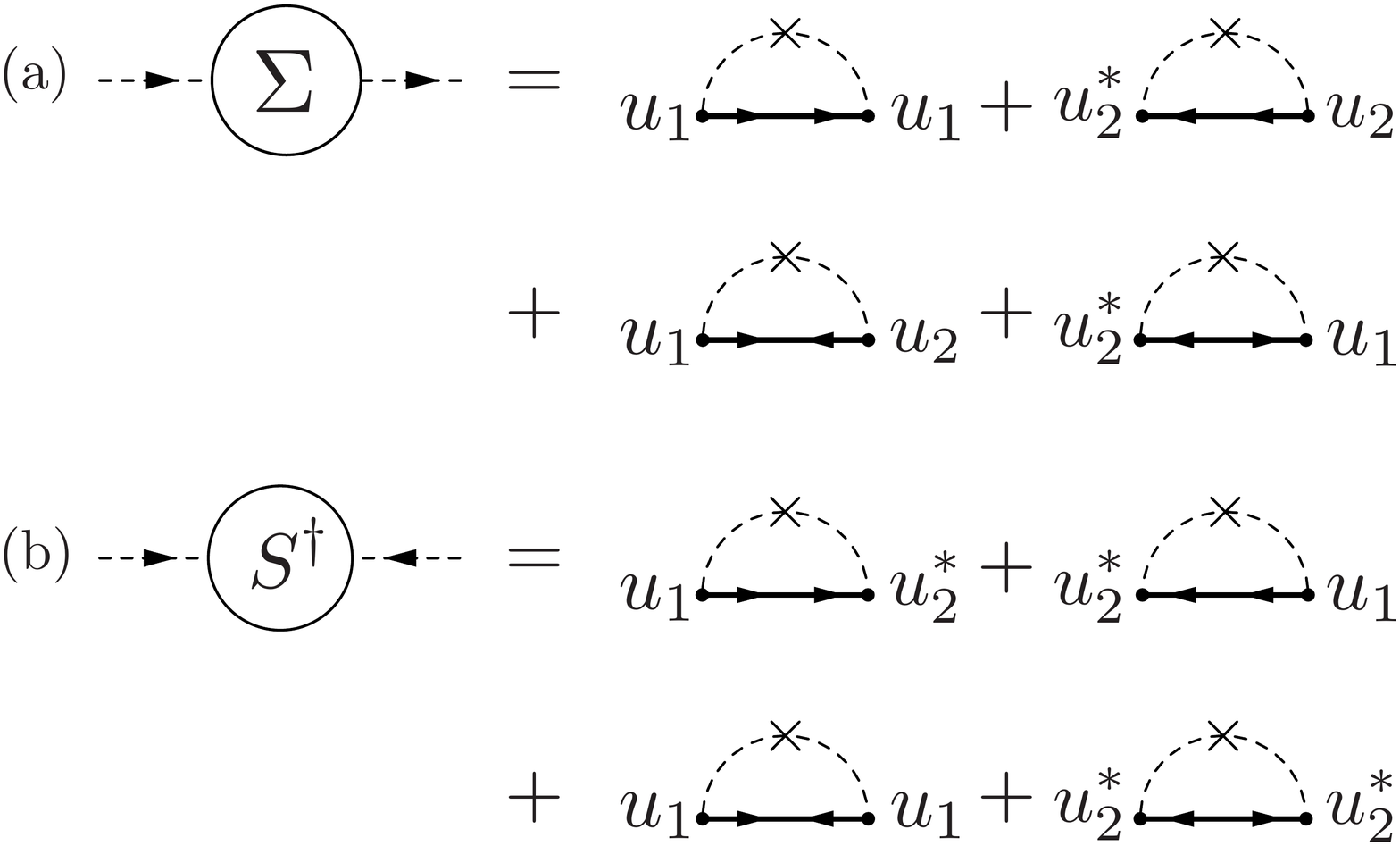}
\caption{
Feynman diagrams relevant to the self-consistent Born approximation.
}
\label{fig:SCBA}
\end{center}
\end{figure}

We also consider the self-energy with self-consistent treatment.
Namely, we consider the diagrams for the self-energy shown in Fig.~\ref{fig:SCBA}.
The self-consistent equations are derived as
\begin{align}
  &\Sigma_{\bm{k}}(\imu\varepsilon_n) = n_{\mathrm{imp}} \int \frac{\diff\bm{q}}{(2\pi)^3} \Big[|u_1(\bm{q} , \bm{k} - \bm{q})|^2 G_{\bm{q}}(\imu\varepsilon_n) \notag \\
  &- 4|u_2(\bm{q} , \bm{k} - \bm{q})|^2 G_{\bm{q}}(-\imu\varepsilon_n) \notag \\
  &- 2u_1(\bm{q} , \bm{k} - \bm{q}) u_2^\ast(\bm{q} , \bm{k} - \bm{q}) F_{\bm{q}}(-\imu\varepsilon_n) \notag \\
  &- 2u_1^\ast(\bm{q} , \bm{k} - \bm{q}) u_2(\bm{q} , \bm{k} - \bm{q}) F^\dagger_{\bm{q}}(-\imu\varepsilon_n) \Big], \\
  &S^\dagger_{\bm{k}}(\imu\varepsilon_n) = n_{\mathrm{imp}} \int \frac{\diff\bm{q}}{(2\pi)^3} \Big[2u_1^\ast(\bm{q} , \bm{k} - \bm{q}) u_2^\ast(\bm{q} , \bm{k} - \bm{q}) \notag \\
  &\times \big( G_{\bm{q}}(\imu\varepsilon_n) - G_{\bm{q}}(-\imu\varepsilon_n) \big) \notag \\
  &+ \frac{1}{2} u_1^\ast(\bm{q} , \bm{k} - \bm{q}) u_1^\ast(\bm{q} , \bm{k} - \bm{q}) \big(F^\dagger_{\bm{q}}(\imu\varepsilon_n) - F^\dagger_{\bm{q}}(-\imu\varepsilon_n) \big) \notag \\
  &+ 2u_2^\ast(\bm{q} , \bm{k} - \bm{q}) u_2^\ast(\bm{q} , \bm{k} - \bm{q}) \big(F_{\bm{q}}(\imu\varepsilon_n) - F_{\bm{q}}(-\imu\varepsilon_n) \big) \Big]. \notag \\ \label{eq:self_consistent_2}
\end{align}
Obviously, these equations reduce to those in the Born approximation in the last subsection if one drops the self-energies in the right-hand side.
In order to search for the concrete solutions, we simplify the equation by replacing the coefficients by the wavevector independent ones, and then the self-energies are also $\bm k$-independent.
We have considered all the parameter space within this approximation, and always have found physical solutions.
Defining the energy-dependent functions $\Gamma_{1,2}(\imu\ep_n)$ by $\Sigma (\imu\ep_n) = - \imu \Gamma_1 (\imu \ep_n)$ and $S^\dg (\imu\ep_n) = - \imu \Gamma_2 (\imu \ep_n)$, we obtain the equations
\begin{align}
  &\Gamma_1(\imu\ep_n) = \frac{\imu\pi n_{\mathrm{imp}} D_0}{\sqrt{-(\varepsilon_n + \mathrm{Re}\Gamma_1(\imu\varepsilon_n))^2 + |\Gamma_2(\imu\varepsilon_n)|^2}} \notag \\
  &\times \Big[\big( |u_1|^2 + 4 |u_2|^2 \big) (\varepsilon_n + \mathrm{Re}\Gamma_1(\imu\varepsilon_n)) \notag \\
  &- 2 u_1 u_2^\ast \Gamma_2^\ast(i\varepsilon_n) - 2 u_1^\ast u_2 \Gamma_2(\imu\varepsilon_n) \Big]
  \label{eq:g1} \\
  &\Gamma_2(\imu\ep_n) = \frac{\imu\pi n_{\mathrm{imp}} D_0}{\sqrt{-(\varepsilon_n + \mathrm{Re}\Gamma_1(\imu\varepsilon_n))^2 + |\Gamma_2(\imu\varepsilon_n)|^2}} \notag \\
  &\times\Big[4 u_1^\ast u_2^\ast (\varepsilon_n + \mathrm{Re}\Gamma_1(\imu\varepsilon_n)) \notag \\
  &- u_1^\ast u_1^\ast \Gamma_2(i\varepsilon_n) - 4 u_2^\ast u_2^\ast \Gamma_2^\ast(\imu\varepsilon_n) \Big].
  \label{eq:g2}
\end{align}
Performing a suitable transformation for the dimensionless expressions of the equations, we find that the equations are controlled by the complex parameter $u_1/|u_2|$.
We note that the solution is identical to the Born approximation discussed in the last subsection at high frequencies.

\begin{figure}[t]
\begin{center}
\includegraphics[width=85mm]{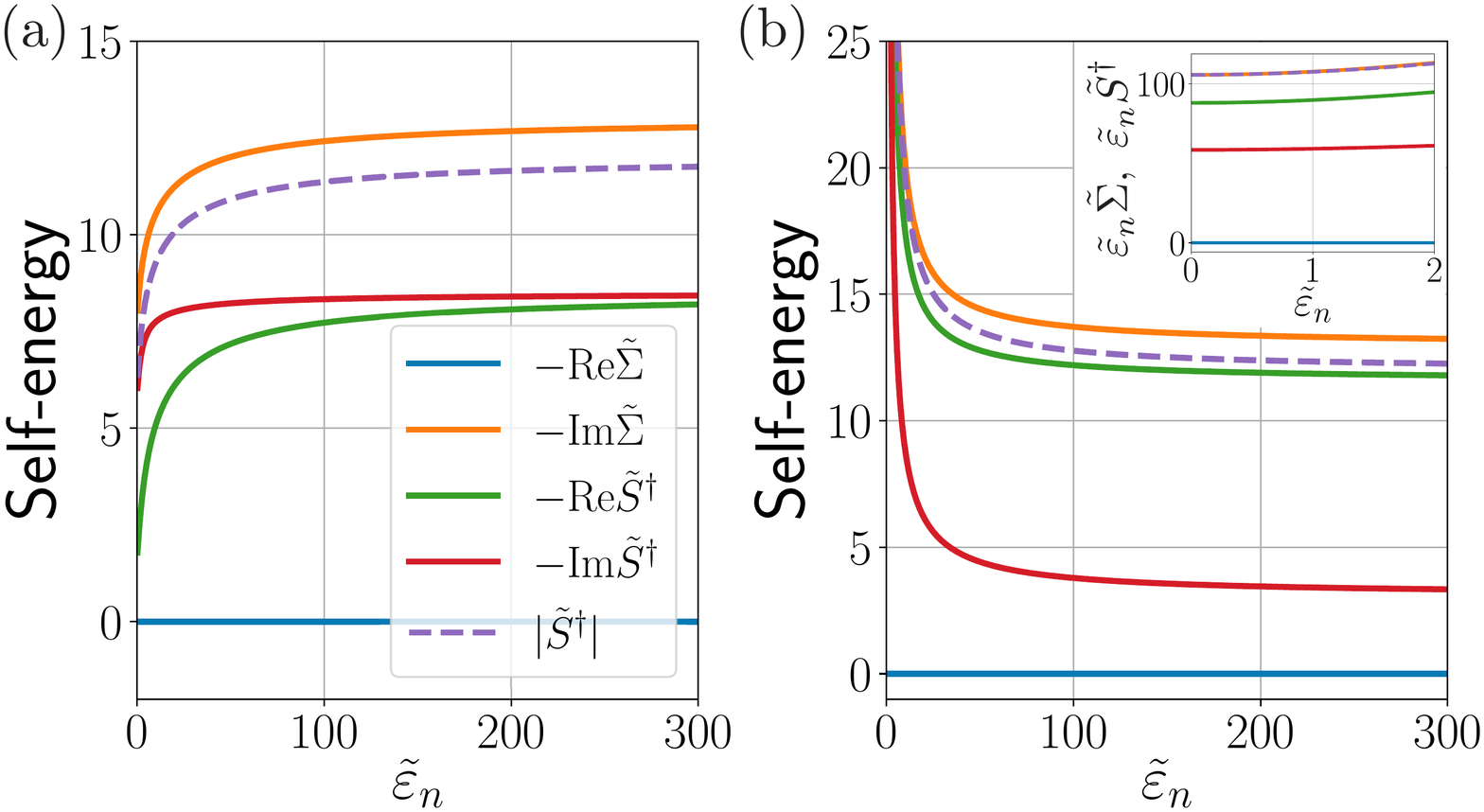}
\caption{
Examples of the solutions of Eqs.~\eqref{eq:g1} and \eqref{eq:g2}. We define dimensionless self-energies and frequency as $\tilde{\Sigma} = \Sigma/(\pi n_{\mathrm{imp}} D_0 |u_2|^2) , \tilde{S}^\dagger = S^\dagger/(\pi n_{\mathrm{imp}} D_0 |u_2|^2) , \tilde{\varepsilon}_n = \varepsilon_n/(\pi n_{\mathrm{imp}} D_0 |u_2|^2)$.
The parameters are chosen as (a) $u_1/|u_2| = 3\exp(\imu \pi/6) , \arg u_2 = \pi/12$ and (b) $u_1/|u_2| = 3\exp(\imu\pi/3) , \arg u_2 = \pi/12$.
The inset shows the self-energy with $\ep_n$ multiplied, indicating $-\ep_n \imag \Sigma(\imu \ep_n) = \ep_n |S^\dg(\imu \ep_n)|$ for $\ep_n\to 0$.
}
\label{fig:solutions}
\end{center}
\end{figure}

We can solve the above simultaneous equations at each frequency.
The calculation examples are shown in Fig.~\ref{fig:solutions}, where we have the two kinds of solutions depending on the parameters both of which are odd function in frequency or imaginary time.
The characteristic behaviors are seen in the low-energy limit $\ep_n \to 0$.
The first kind has the form
\begin{align}
\Sigma(\imu\ep_n \to 0) &= - \imu \Gamma_1 \sgn\ep_n,
\\
S^\dg (\imu\ep_n \to 0) &= - \imu \Gamma_2 \sgn\ep_n,
\end{align}
which is similar to the results in the simple Born approximation.
We can explicitly obtain the solution as
\begin{align}
  \frac{\Gamma_2}{\Gamma_1} =
  \begin{cases}
    \displaystyle \frac{2u_2^\ast}{\mathrm{Re}u_1} & (2 < |\mathrm{Re}u_1|/|u_2|),\vspace{5pt}\\
    \displaystyle \frac{\mathrm{Re}u_1}{2u_2} &(\mathrm{Im}u_1/|u_2| = 0 \text{ and } 0 < |\mathrm{Re}u_1|/|u_2| < 2)
  \end{cases}
\end{align}
for $|\real u_1|/|u_2| > 2$ or $\imag u_1/|u_2| =0$.
The calculation example is shown in Fig.~\ref{fig:solutions}(a).

On the other hand, the solution of the second kind is obtained for $|\real u_1|/|u_2| <2$ and $\imag u_1/|u_2| \neq 0$, which has the frequency dependence
\begin{align}
  \Sigma(\imu\ep_n \to 0) &= \frac{|V|^2}{\imu \ep_n},
  \\
  S^\dg(\imu\ep_n \to 0)  &= \frac{V^2}{\imu \ep_n},
\end{align}
where $V$ is a complex constant.
The calculation example is shown in Fig.~\ref{fig:solutions}(b).
The absolute value $|V|$ cannot be written in the simple analytic form, but is determined easily from the numerical calculations.

\subsection*{3. Single-particle spectral functions}

From the normal and anomalous Green functions, we obtain several physical quantities of interest.
The first one is the single-particle spectral function and density of states defined by
\begin{align}
  A_{\bm k}(\varepsilon) &= -\frac{1}{\pi} \imag G_{\bm k11}(\varepsilon+\imu\eta),
  \\
  D(\varepsilon) &= \sum_{\bm k} A_{\bm k}(\varepsilon)
  \simeq D_0 \int \diff \xi_{\bm k} A_{\bm k}(\varepsilon).
\end{align}
The momentum distribution function is given by
\begin{align}
  n_{\bm k} &= \la \al^\dg_{\bm k} \al_{\bm k} \ra.
\end{align}
In addition, the time-dependent pair amplitude can also be calculated.
Whereas the static quantity such as $\la \al_{\bm k } \al_{-\bm k}\ra$ is zero due to the inversion symmetry, the pair amplitude is finite at different times.
Specifically we consider the spatially local quantity
\begin{align}
  F(\tau) &= T \sum_{n,\bm k} F_{\bm k} (\imu\ep_n) \epn^{-\imu\ep_n \tau}.
\end{align}
The above quantities are graphically shown in the main text.

The real time dynamics is also obtained by performing the analytic continuation.
The retarded Green function is given by
\begin{align}
  G^{\rm R}(t) = - \imu \theta (t) \la \{ \al_{\bm k}(t), \al_{\bm k}^\dg \} \ra,
\end{align}
where $\theta(t)$ is the step function.
Here we have considered the real-time Heisenberg picture $A(t) = \epn^{\imu t\mathscr H} A \epn^{-\imu t\mathscr H}$.
In the following, we summarize the above physical quantities for the solutions of both the first and second kinds.

\subsubsection*{{\rm (i)} Solution of the first kind}
We consider the self-energies
\begin{align}
  \Sigma(\imu\ep_n) = -\imu\Gamma_1 \sgn\ep_n
  ,\ \
  S^\dagger(\imu\ep_n) = -\imu\Gamma_2 \sgn\ep_n
  ,
\end{align}
with $\Gamma_1 > |\Gamma_2|$. The density of states is
\begin{align}
  \frac{D(\ep)}{D_0} &= \frac{|\ep| + \imu\Gamma_1}{2\sqrt{(|\ep| + \imu\Gamma_1)^2 + |\Gamma_2|^2}}
  + {\rm c.c.}
\end{align}
The momentum distribution function is
\begin{align}
  n_{\bm k} &= \frac{\xi_{\bm k}}{2\pi \sqrt{|\Gamma_2|^2- \xi_{\bm k}^2}}
  \ln \left(
  \frac{ \Gamma_1 - \sqrt{|\Gamma_2|^2- \xi_{\bm k}^2} }
  { \Gamma_1 +  \sqrt{|\Gamma_2|^2- \xi_{\bm k}^2} }
  \right) + \frac 1 2
\end{align}
at zero temperature.
The spatially local pair amplitude at $T=0$ is
\begin{align}
  \frac{F^\dagger(\tau)}{\Gamma_2} &=  \int _0 ^\infty \diff \ep
  \left[
  \frac{\imu}{ 2\sqrt{(|\ep| + \imu\Gamma_1)^2 + |\Gamma_2|^2}} + {\rm c.c.}
  \right] \epn^{-\ep \tau}
\end{align}
for $\tau>0$, which is numerically integrated.

The real time dependence of the retarded Green function ($t>0$) is given by
\begin{align}
  G^{\rm R}_{\bm k}(t) &=
  - \frac{\imu}{2} \left(
  1 + \frac{\xi_{\bm k}}{\sqrt{\xi_{\bm k}^2 - |\Gamma_2|^2}}
  \right)
  \epn^{- \imu \sqrt{\xi_{\bm k}^2 - |\Gamma_2|^2} \, t - \Gamma_1 t} \notag \\
  & - \frac{\imu}{2} \left(
  1 - \frac{\xi_{\bm k}}{\sqrt{\xi_{\bm k}^2 - |\Gamma_2|^2}}
  \right)
  \epn^{ \imu \sqrt{\xi_{\bm k}^2 - |\Gamma_2|^2} \, t - \Gamma_1 t}
  \end{align}
  for $|\xi_{\bm k}| > |\Gamma_2|$ and
  \begin{align}
  G^{\rm R}_{\bm k}(t) &=
  - \frac{1}{2} \left(
  \imu + \frac{\xi_{\bm k}}{\sqrt{|\Gamma_2|^2 - \xi_{\bm k}^2}}
  \right)
  \epn^{- (\Gamma_1 - \sqrt{|\Gamma_2|^2 - \xi_{\bm k}^2}) t}
  \nonumber \\
  &- \frac{1}{2} \left(
  \imu - \frac{\xi_{\bm k}}{\sqrt{|\Gamma_2|^2 - \xi_{\bm k}^2}}
  \right)
  \epn^{- (\Gamma_1 + \sqrt{|\Gamma_2|^2 - \xi_{\bm k}^2}) t}
\end{align}
for $|\xi_{\bm k}| < |\Gamma_2|$.

\subsubsection*{{\rm (ii)} Solution of the second kind}

We summarize the results for the self-energies
\begin{align}
  \Sigma(\imu\ep_n) = |V|^2 / \imu\ep_n
  ,\ \
  S^\dagger (\imu\ep_n) = V^2 / \imu\ep_n.
  \end{align}
  The density of states is
  \begin{align}
  \frac{D(\ep)}{D_0} &= \frac{\pi }{\sqrt 2} |V| \delta(\ep)
  + \frac{\ep^2 - |V|^2}{\ep\sqrt{\ep^2 - 2|V|^2}} \theta(\ep - \sqrt 2 |V|).
  \end{align}
  The momentum distribution function is
  \begin{align}
  n_{\bm k} &= \frac{E_{\bm k}-\xi_{\bm k}[1- 2f(E_{\bm k})]}{2E_{\bm k}},
  \end{align}
  where $E_{\bm k} = \sqrt{\xi_{\bm k}^2 + 2|V|^2}$ and $f(x) = 1 / (\epn^{x/T}+1)$.
  The spatially local pair amplitude is
  \begin{align}
  \frac{F^\dagger(\tau)}{V} &= \frac{\pi V}{2\sqrt{2} |V|} - V \int_{\sqrt 2 |V|}^\infty \diff \ep
  \frac{\epn^{-\ep \tau}}{\ep \sqrt{\ep^2 - 2|V|^2}}
  \end{align}
  for $\tau>0$,
  which is numerically integrated.
  The real time dependence of the Green function is given by
  \begin{align}
  G^{\rm R}_{\bm k}(t) &= - \frac{\imu}{2E_{\bm k}^2}
  \Big[
  2|V|^2 + (\xi_{\bm k}^2+|V|^2 + \xi_{\bm k}E_{\bm k})\epn^{-\imu E_{\bm k}t}
  \nonumber \\
  &
  + (\xi_{\bm k}^2+|V|^2 - \xi_{\bm k}E_{\bm k})\epn^{\imu E_{\bm k}t}
  \Big]
\end{align}
for $t>0$.

\section*{SM B. Interaction with bosons}

The interaction Hamiltonian is defined as Eq~\eqref{eq:imp} and Eq~\eqref{eq:int_boson} in the main text.
With the Hermiticity and inversion symmetry, we obtain the relations similar to Eqs.~(\ref{eq:hermite_1}--\ref{eq:inversion}) for $g_{1,2}(\bm k,\bm q)$.
The derivation of the self-energies is similar to the standard procedure used in the Fermi liquid theory \cite{AGD_book}.

\begin{figure}[t]
\begin{center}
\includegraphics[width=85mm]{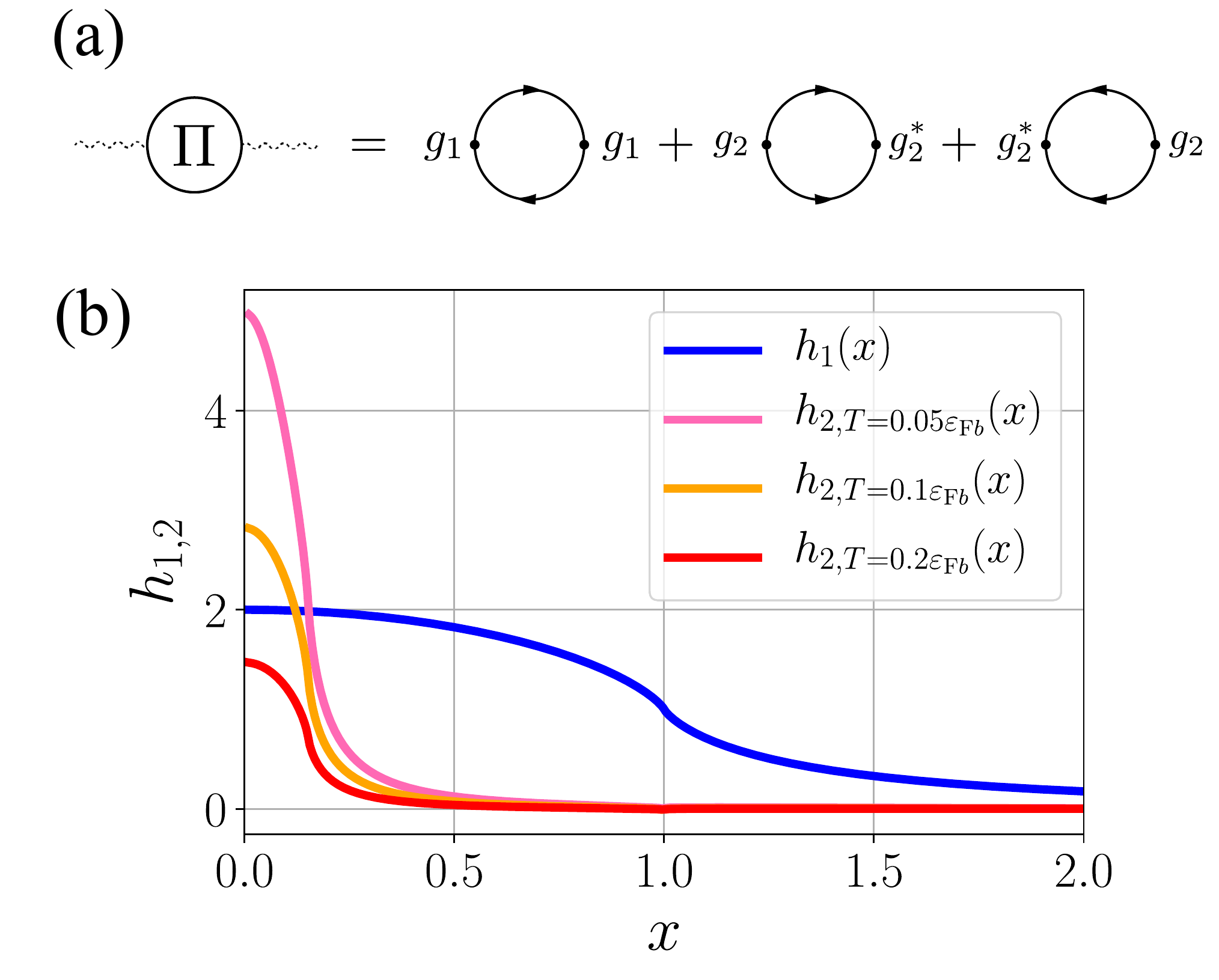}
\caption{
(a) Self-energy for bosons.
(b) Wavevector dependence of $h_{1,2}$ defined in Eqs.~\eqref{eq:h_1} and \eqref{eq:h_2} at several temperatures.
The cutoff wavevector is chosen as $\omega_{\mathrm{C}}/\varepsilon_{\mathrm{F}b} = 0.3$
}
\label{fig:functions}
\end{center}
\end{figure}

\subsection*{1. Boson self-energy}
We first derive the boson self-energy.
The Green function of boson is defined by
\begin{align}
  D_{\bm{q}}(\tau) = - \langle \mathcal{T} \phi_{\bm{q}}(\tau) \phi_{-\bm{q}} \rangle, \\
  \phi_{\bm{q}} = \imu\sqrt{\frac{\omega_{0,\bm{q}}}{2}} (b_{\bm{q}} - b_{-\bm{q}}^\dagger).
\end{align}
Fourier transformation is given by
\begin{align}
  D_{\bm q}(\imu\omega_n) &= \int_0^{1/T} \diff \tau D_{\bm q} (\tau) \epn^{\imu\omega_n \tau},
\end{align}
where $\omega_n = 2n\pi T$ is the bosonic Matsubara frequency.
The self-energy $\Pi_{\bm{q}}(\imu\omega_n)$ is defined by
\begin{align}
  D_{\bm{q}}(\imu\omega_n)^{-1} &= D^0_{\bm{q}}(\imu\omega_n)^{-1} - \Pi_{\bm{q}}(\imu\omega_n),
\end{align}
with the zeroth-order Green function $D^0_{\bm{q}}(\imu\omega_n) = \omega_{0,\bm{q}}^2/[(\imu\omega_n)^2 - \omega_{0,\bm{q}}^2]$.
The diagrammatic contributions are illustrated in Fig.~\ref{fig:functions}(a), and are then given as follows:
\begin{align}
  &\Pi_{\bm{q}}(\imu\omega_n) = \Pi_{1,\bm{q}}(\imu\omega_n) + 2\Pi_{2,\bm{q}}(\imu\omega_n), \\
  &\Pi_{1,\bm{q}}(\imu\omega_n) = \frac{T}{(2\pi)^3} \sum_{m} \int \diff\bm{k} \, |g_1(\bm{k} , -\bm{q})|^2 \notag \\
  &\quad \times G^0_{\bm{k}}(\imu\varepsilon_m) G^0_{\bm{k} - \bm{q}}(\imu\varepsilon_m - \imu\omega_n), \\
  &\Pi_{2,\bm{q}}(\imu\omega_n) \notag = -\frac{T}{(2\pi)^3} \sum_{m} \int \diff\bm{k} |g_2(\bm{k} , -\bm{q})|^2
  G^0_{\bm{k}}(\imu\varepsilon_m) \notag \\
  &\quad \times[G^0_{-\bm{k} + \bm{q}}(-\imu\varepsilon_m + \imu\omega_n) + G^0_{-\bm{k} + \bm{q}}(-\imu\varepsilon_m - \imu\omega_n)],
\end{align}
where $G^0_{\bm{k}}(\imu \varepsilon_n)$ is a free particle Green function of bogolon.
First, we find an explicit form of $\Pi_{1,\bm{q}}(\imu \omega_n)$.
For the concrete calculation, we introduce the wavevector-averaged coupling constant $\overline{\Gamma}_{1,\bm{q}} \equiv \langle |g_1(\bm{k} , -\bm{q})|^2 \rangle_{\bm{k}}$.
Then we obtain the real part of retarded self-energy at low $\omega$ as follows:
\begin{align}
  \mathrm{Re}\Pi^{\mathrm{R}}_{1,\bm{q}} (\omega) = -\frac{\overline{\Gamma}_{1,\bm{q}} m_b k_{\mathrm{F}b}}{(2\pi)^2}h_1\left(\frac{q}{2k_{\mathrm{F}b}}\right),
\end{align}
where
\begin{align}
  h_1(x) \equiv 1 + \frac{1 - x^2}{2x} \ln\left| \frac{1 + x}{1 - x} \right| \label{eq:h_1}
\end{align}
is the Lindhard function shown in Fig.~\ref{fig:functions}(b).
We also get the imaginary part as
\begin{align}
  \mathrm{Im}\Pi^{\mathrm{R}}_{1,\bm{q}} (\omega)
  = -\frac{\overline{\Gamma}_{1,\bm{q}} m_b^2}{2\pi q}\omega \theta(2k_{\mathrm{F}b} - q).
\end{align}
These are the same form as those for the usual electrons.
On the other hand, in order to calculate $\Pi^{\mathrm{R}}_{2,\bm{q}}(\omega)$, we introduce the wavevector-averaged coupling constant $\overline{\Gamma}_{2,\bm{q}} \equiv \langle |g_2(\bm{k} , -\bm{q})|^2 \rangle_{\bm{k}}$. The result of the calculation of $\Pi^{\mathrm{R}}_{2,\bm{q}}(\omega)$ at low $\omega$ is as follows:
\begin{align}
  \mathrm{Re}\Pi^{\mathrm{R}}_{2,\bm{q}}(\omega)
  &= - \frac{\overline{\Gamma}_{2,\bm{q}} m_b k_{\mathrm{F}b}}{(2\pi)^2} h_{2,T}\left(\frac{q}{2k_{\mathrm{F}b}}\right), \\
  \mathrm{Im}\Pi^{\mathrm{R}}_{2,\bm{q}} (\omega)
  &= -\frac{\overline{\Gamma}_{2,\bm{q}} m_b^2}{2\pi q}\omega \theta\big((1 + \sqrt{3})k_{\mathrm{F}b} - q\big),
\end{align}
where
\begin{align}
  &h_{2,T}(x) \equiv \notag \\
  & \int_{\sqrt{1 - \omega_{\mathrm{C}}/\varepsilon_{\mathrm{F}b}}}^{\sqrt{1 + \omega_{\mathrm{C}}/\varepsilon_{\mathrm{F}b}}}
\hspace{-2mm}
\diff y \frac{y}{x} \tanh\left(\frac{y^2 - 1}{2T/\varepsilon_{\mathrm{F}b}}\right) \ln\left| \frac{y^2 - 1 + 2x^2 + 2xy}{y^2 - 1 + 2x^2 - 2xy} \right|, \label{eq:h_2}
\end{align}
where the energy integration is performed within the cutoff frequency of bosons $\omega_{\mathrm{C}}$.
We have assumed that the dominant contributions enter at small $q$.
Note that $h_{2,T}(q/2k_{\mathrm{F}b})$ diverges logarithmically in the infrared regime at zero temperature limit.
This corresponds to the Cooper instability intrinsic to Fermi surfaces.
However, this divergence is suppressed at finite $T$ as shown in Fig.~\ref{fig:functions}(b), in which we are interested.

For later discussion, we derive the explicit form of boson Green function. The Dyson equation is given as follows:
\begin{align}
  [D^{\mathrm{R}}_{\bm{q}}(\omega)]^{-1} = \frac{1}{\omega_{0,\bm{q}}^2} \left( \omega^2 - \omega_{\bm{q}}^2 - 2\imu\omega \gamma_{\bm{q}} \right),
\end{align}
where
\begin{align}
  \omega_{\bm{q}} &= \omega_{0,\bm{q}}\sqrt{1 - \eta_{\bm{q}}}, \\
  \eta_{\bm{q}} &= \frac{\overline{\Gamma}_{1,\bm{q}} m_b k_{\mathrm{F}b}}{(2\pi)^2}h_1\left(\frac{q}{2k_{\mathrm{F}b}}\right) + 2\frac{\overline{\Gamma}_{2,\bm{q}} m_b k_{\mathrm{F}b}}{(2\pi)^2}h_{2,T}\left(\frac{q}{2k_{\mathrm{F}b}}\right), \\
  \gamma_{\bm{q}} &= \frac{\overline{\Gamma}_{1,\bm{q}} m_b^2 \omega_{0,\bm{q}}^2}{4\pi q} \theta(2k_{\mathrm{F}b} - q) \notag \\
  &\quad + \frac{\overline{\Gamma}_{2,\bm{q}} m_b^2 \omega_{0,\bm{q}}^2}{2\pi q} \theta\big((1 + \sqrt{3})k_{\mathrm{F}b} - q\big).
\end{align}
Therefore, we get the explicit form of the Green function as
\begin{align}
  D_{\bm{q}}^{\mathrm{R}}(\omega)
  = \frac{\omega_{0,\bm{q}}^2}{2\omega_{\bm{q}}} \left( \frac{1}{\omega - \omega_{\bm{q}} + \imu\gamma_{\bm{q}}} - \frac{1}{\omega + \omega_{\bm{q}} + \imu\gamma_{\bm{q}}} \right).
\end{align}
which will be used to obtain the bogolon self-energies.

\subsection*{2. Bogolon normal self-energy}

Next, we derive the bogolon normal self-energy given in Eq.~\eqref{eq:G}. The diagram of the second-order self-energy is shown in Fig.~\ref{fig:diagram}(b) of the main text.
The corresponding self-energy $\Sigma_{\bm{k}}(\imu\varepsilon_n)$ is given as follows:
\begin{align}
  \Sigma_{\bm{k}}(\imu\varepsilon_n) &= \Sigma_{1,\bm{k}} (\imu\varepsilon_n) + 2 \Sigma_{2,\bm{k}} (\imu\varepsilon_n), \\
  \Sigma_{1,\bm{k}}(\imu\varepsilon_n)
  &= -\frac{T}{(2\pi)^3} \sum_{m} \int \diff\bm{k}' \, |g_1(\bm{k}' , \bm{k} - \bm{k}')|^2 \notag \\
  &\quad\times G^0_{\bm{k}'} (\imu\varepsilon_m) D_{\bm{k} - \bm{k}'} (\imu\varepsilon_n - \imu\varepsilon_m), \\
  \Sigma_{2,\bm{k}}(\imu\varepsilon_n)
  &= \frac{2T}{(2\pi)^3} \sum_{m} \int \diff\bm{k}' \, |g_2(\bm{k}' , \bm{k} - \bm{k}')|^2 \notag \\
  &\quad \times G^0_{\bm{k}'}(\imu\varepsilon_m) D_{\bm{k} - \bm{k}'} (\imu\varepsilon_n + \imu\varepsilon_m).
\end{align}
We calculate $\Sigma_{1,\bm{k}}(\imu\varepsilon_n)$ at first.
For the evaluation of $\Sigma_{\bm{k}}(\imu\varepsilon_n)$, we replace the Matsubara sums by the energy integral.
By using the spectral representation for the non-interacting Green function
\begin{align}
  G^{0,\mathrm{R}}(\bm{k} , \varepsilon) = \int_{-\infty}^\infty \diff\varepsilon' \frac{A_{\bm{k}}^0(\varepsilon')}{\varepsilon + \imu\eta - \varepsilon'},
\end{align}
the retarded self-energy has the form
\begin{align}
  &\Sigma^{\mathrm{R}}_{1,\bm{k}}(\varepsilon)
  = \frac{1}{(2\pi)^4} \int \diff\bm{k}' \int_{-\infty}^\infty \diff\varepsilon' \int_{-\infty}^\infty \diff\omega \notag \\
  &\quad \times |g_1(\bm{k}' , \bm{k} - \bm{k}')|^2 \frac{A_{\bm{k}'}^0(\varepsilon') \mathrm{Im}D^{\mathrm{R}}_{\bm{k} - \bm{k}'}(\omega)}{\omega + \varepsilon' - \varepsilon - \imu\eta} \notag \\
  &\quad \times \left[ \tanh\left(\frac{\varepsilon'}{2T}\right) + \coth\left(\frac{\omega}{2T}\right) \right].
\end{align}
To evaluate the integral, we again introduce the wavevector-averaged coupling constant $\overline{\Gamma}_{3,\bm{k}} \equiv \langle |g_1(\bm{k}' , \bm{k} - \bm{k}')|^2 \rangle_{\bm{k}'} \in \mathbb R$.
Using the  variable transformation $\bm{k}' \to (q , \xi_{\bm{k}'})$ with $\bm{q} = \bm{k} - \bm{k}'$, we obtain
\begin{align}
  &\Sigma^{\mathrm{R}}_{1,\bm{k}}(\varepsilon)
  = \frac{\overline{\Gamma}_{3,\bm{k}} m_b}{(2\pi)^3 k} \int_0^{k_1} \diff q\, q \int_{-\infty}^\infty \diff\omega \int_{\xi_{|\bm{k}| - q}}^{\xi_{|\bm{k}| + q}} \diff\varepsilon' \notag \\
  &\quad \times \frac{\mathrm{Im}D^{\mathrm{R}}_{\bm{q}}(\omega)}{\omega + \varepsilon' - \varepsilon - \imu\eta} \left[ \tanh\left(\frac{\varepsilon'}{2T}\right) + \coth\left(\frac{\omega}{2T}\right) \right],
\end{align}
where $k_{1} = \mathrm{min}\{ q_{\mathrm{C}} , 2k_{\mathrm{F}b} \}$, and $q_{\mathrm{C}}$ is a wavevector cutoff of bosons.
We can approximate $\xi_{|\bm{k}| - q}$ and $\xi_{|\bm{k}| + q}$ by $-\infty$ and $+\infty$, respectively \cite{AGD_book}.
This procedure is checked for $\ep_{\mathrm{Fb}}/\omega_{\rm C} \gg 1$ by comparing the result with the numerical calculation without using the extrapolation, regardless of the cutoff wavevector $k_1$.
Moreover, we use the relation $\mathrm{Re}D^{\mathrm{R}}_{\bm{q}}(\omega) = \omega_{0,\bm{q}}^2/(\omega^2 - \omega_{\bm{q}}^2)$.
Then, the real part of $\Sigma_{1,\bm{k}}^{\mathrm{R}}(\varepsilon)$ is given as follows:
\begin{align}
  \mathrm{Re}\Sigma^{\mathrm{R}}_{1,\bm{k}}(\varepsilon)
  &= - \frac{\overline{\Gamma}_{3,\bm k} \zeta}{\omega^*} \ep
\end{align}
for $\varepsilon \ll \omega_{\mathrm{C}}$, where we have introduced the quantities
\begin{align}
  \omega^* &= \frac{
  \displaystyle
  \int_0^{k_1} \diff q\, q \frac{\omega_{0,\bm{q}}^2}{\omega_{\bm{q}}^2}
  }{
  \displaystyle
  \int_0^{k_1} \diff q\, q \frac{\omega_{0,\bm{q}}^2 \gamma_{\bm{q}}}{\omega_{\bm{q}}^4}
  }
  ,
  \\
  \frac{\zeta}{\omega^{*}} &= \frac{m_b}{4\pi^2 k_{\mathrm{F}b}}
  \int_0^{k_1} \diff q\, q \frac{\omega_{0,\bm{q}}^2}{\omega_{\bm{q}}^2}
\end{align}
to make the notations simple.
With $\mathrm{Im}D^{\mathrm{R}}_{\bm{q}}(\omega) \simeq -2\omega_{0,\bm{q}}^2 \gamma_{\bm{q}} \omega / \omega_{\bm{q}}^4$ for $\omega \ll \gamma_{\bm{q}}$, the imaginary part of $\Sigma_{1,\bm{k}}^{\mathrm{R}}(\varepsilon)$ is obtained as
\begin{align}
  \mathrm{Im}\Sigma^{\mathrm{R}}_{1,\bm{k}}(\varepsilon)
  &= - \frac{\overline{\Gamma}_{3,\bm k} \zeta}{\omega^{*2}} (\pi^2 T^2 + \varepsilon^2).
\end{align}
The other contribution can also be evaluated with a similar procedure.
We list the result below:
\begin{align}
  &\Sigma^{\mathrm{R}}_{2,\bm{k}} (\varepsilon)
= - \frac{2\overline{\Gamma}_{4,\bm k} \zeta}{\omega^*} \ep
- \imu \frac{2\overline{\Gamma}_{4,\bm k} \zeta}{\omega^{*2}} (\pi^2 T^2 + \varepsilon^2),
\end{align}
where $\overline{\Gamma}_{4,\bm{k}} = \langle |g_2(\bm{k}' , \bm{k} - \bm{k}')|^2 \rangle_{\bm{k}'} \in \mathbb R$.

\subsection*{3. Bogolon anomalous self-energy}
In this subsection, we derive the bogolon anomalous self-energy.
The diagram of the second-order self-energy is shown in Fig.\ref{fig:diagram} (c) of the main text. With inversion symmetry, the corresponding self-energy $S_{\bm{k}}^\dagger(\imu\varepsilon_n)$ is given as follows:
\begin{align}
  &S^\dagger_{\bm{k}}(\imu\varepsilon_n) = -\frac{2T}{(2\pi)^3} \sum_{m} \int \diff\bm{k}' \, g_1^\ast(\bm{k}' , \bm{k} - \bm{k}') g_2^\ast(\bm{k}' , \bm{k} - \bm{k}') \notag \\
  &\quad \times G^0_{\bm{k}'} (\imu\varepsilon_m) [D_{\bm{k} - \bm{k}'} (-\imu\varepsilon_n + \imu\varepsilon_m) - D_{\bm{k} - \bm{k}'} (\imu\varepsilon_n + \imu\varepsilon_m)].
\end{align}
The explicit form of the retarded self-energy $S_{\bm{k}}^{\dagger \mathrm{R}}(\varepsilon)$ can be derived by a similar manner to the previous subsection.
The result is written as
\begin{align}
  &S_{\bm{k}}^{\dagger \mathrm{R}}(\varepsilon) =
  - \frac{4\overline{\Gamma}_{5,\bm k} \zeta}{\omega^*} \ep
  - \imu \frac{4\overline{\Gamma}_{5,\bm k} \zeta}{\omega^{*2}} (\pi^2 T^2 + \varepsilon^2),
\end{align}
where $\overline{\Gamma}_{5,\bm{k}} = \langle g_1^\ast(\bm{k}' , \bm{k} - \bm{k}') g_2^\ast (\bm{k}' , \bm{k} - \bm{k}') \rangle_{\bm{k}'}$.

\subsection*{4. Analytic continuation}

For the characterization of the frequency-dependent functional forms, it is suitable to see the physical quantities on the imaginary (or Matsubara) axis.
The results on the retarded bogolon self-energies obtained in the previous subsections can be analytically continued to the imaginary axis as
\begin{align}
  \Sigma_{\bm{k}} (\imu\ep_n) &= a_{\bm{k}}\imu\varepsilon_n + \imu b_{\bm{k}}\left[ \pi^2 T^2 + (\imu\varepsilon_n)^2 \right] \sgn \varepsilon_n,
  \\
  S_{\bm{k}}^\dg (\imu\ep_n) &= c_{\bm{k}}\imu\varepsilon_n + \imu d_{\bm{k}}\left[ \pi^2 T^2 + (\imu\varepsilon_n)^2 \right] \sgn \varepsilon_n,
\end{align}
where
\begin{align}
  a_{\bm{k}} &= -(\overline{\Gamma}_{3,\bm{k}} + 4\overline{\Gamma}_{4,\bm{k}}) \frac{\zeta}{\omega^*},
  \\
  b_{\bm{k}} &= \frac{a_{\bm{k}}}{\omega^\ast}
  , \\
  c_{\bm{k}} &= - 4 \overline{\Gamma}_{5,\bm{k}} \frac{\zeta}{\omega^*}
  , \\
  d_{\bm{k}} &= \frac{c_{\bm{k}}}{\omega^\ast},
\end{align}
from which we can construct both the retarded and advanced Green functions.
Note that $a_{\bm{k}}$ and $b_{\bm{k}}$ are real, while $c_{\bm{k}}$ and $d_{\bm{k}}$ are complex.

The electron-electron interaction is also expected to lead to a similar effect discussed above.
The imaginary part should be calculated in a manner similar to Ref. \cite{AGD_book},
and then the real part may be reconstructed through the Kramers-Kronig relation \cite{Miyake, Jacko}.

\section*{SM C. Connection to $j=3/2$ fermion model}

\subsection*{1. Electronic Hamiltonian and bogolons}

We consider the $j=3/2$ fermion model as the simplest description for the inversion symmetric Bogoliubov Fermi surfaces \cite{Agterberg17,Tamura20}.
The Hamiltonian is given by
\begin{align}
  &\mathscr H = \sum_{\bm k} \vec c_{\bm k}^\dg \left[
  \Big( \frac{\bm k^2}{2m_e}  - \ep_{{\rm F}e} \Big) \hat 1
  + \beta (\bm k\cdot \hat {\bm J} )^2
  \right] \vec c_{\bm k}
  \nonumber \\
  &+
  \sum_{\bm k} \vec c_{\bm k}^\dg \left[
  \Delta_1 k_z (k_x + \imu k_y)\hat E +
  \frac{2\Delta_0}{\sqrt 3} \lceil \hat J_z (\hat J_x + \imu \hat J_y) \rfloor \hat E
  \right] \vec c_{-\bm k}^{\dg \rm T}
  \nonumber \\
  &+ {\rm H.c.},
  \label{eq:elec_ham}
\end{align}
where $\vec c_{\bm k} = (c_{\bm k, 3/2}, c_{\bm k, 1/2}, c_{\bm k, -1/2}, c_{\bm k, -3/2})^{\rm T}$ is the spin $3/2$ spinor of electrons.
The vector $\hat {\bm J}$, which is $4\times 4$ matrix, represents a spin-3/2 operator (or dipole), and $\hat E$ is the antisymmetric matrix  defined in Ref.~\cite{Tamura20}.
The square bracket symmetrize the product of matrices as $\lceil \hat A \hat B \rfloor = (\hat A \hat B + \hat B \hat A)/2$.

The time-reversal-symmetry broken superconducting state with inversion symmetry generally realizes the Bogoliubov-Fermi surfaces which are topologically protected \cite{Agterberg17}.
Then the resultant effective low-energy Hamiltonian in Eq.~\eqref{eq:ham_zero} of the main text is derived and the bogolon operators are given by
\begin{align}
  \al_{\bm k} &= \sum_m \big(
  u_{\bm{k} , m} c_{\bm{k} , m} + v_{\bm{k} , m} c^\dg_{-\bm{k} , m}
  \big),
\end{align}
where the coefficients $u$ and $v$ are obtained by diagonalizing the Hamiltonian \eqref{eq:elec_ham}.
The odd-frequency pair amplitude of bogolons is now defined by
\begin{align}
  &F_{\bm k}(\tau) = - \la \mathcal T \al_{- \bm k}(\tau) \al_{\bm k} \ra,
\end{align}
which is an odd function with respect to time and is induced from the disorder and  correlation effects.

    \begin{table}[t]
        \begin{tabular}{cc}
\hline
            Rank & $\eta$ \\ \hline
$r=0$ (monopole/singlet) & 1 \\
$r=1$ (dipole/triplet) & $x,y,z$ \\
$r=2$ (quadrupole/quintet) & $xy,yz,zx,z^2,x^2-y^2$ \\
$r=3$ (octupole/septet) &
\begin{tabular}{c}
$xyz, x z^2, y z^2, z^3, z(x^2-y^2),$\\
$x(x^2-3y^2), y(3x^2-y^2)$
\end{tabular}
\\
\hline
      \end{tabular}
        \caption{
List of the types of diagonal and offdiagonal physical quantities classified by the rank $r$.
See Ref.~\cite{Tamura20} for the detailed forms of the matrices $\hat O^\eta$.\label{tab:list}
        }
    \end{table}

Now we consider the time-dependent order parameters in terms of the original electrons as
\begin{align}
  M^{\eta}_{\bm k}(\tau) &= \la \mathcal T \vec c_{\bm k}^\dg \hat O^{\eta} \vec c_{\bm k}(\tau) \ra,
  \\
  P^{\eta}_{\bm k}(\tau) &= \la \mathcal T \vec c_{\bm k}^\dg \hat O^{\eta} \hat E \vec c^{\dg\rm T}_{-\bm k}(\tau) \ra,
  \\
  P^{\dg\eta}_{\bm k}(\tau) &= \la \mathcal T \vec c_{- \bm k}^{\rm T} (\hat O^{\eta} \hat E)^\dg \vec c_{\bm k}(\tau) \ra.
\end{align}
The $4\times 4$ matrices $\hat O^\eta$ are defined in Ref.~\cite{Tamura20}, where all the order parameters are exhausted by this expression.
The pair amplitude corresponds to the multiplet pairs, i.e., electron ($P$) and hole ($P^\dg$) pair amplitudes.
The index $\eta$ represents an identifier for the multipoles and multiplet pairs, and is classified by the rank $r$ which is defined up to $2j$.
(Note that the symbol $\eta$ in this paper corresponds to $\eta'$ in Ref.~\cite{Tamura20}).
For the diagonal quantity, $r=0,1,2,3$ corresponds to one monopole, three dipoles, five quadrupole, and seven octupole.
For the off-diagonal quantity, $r=0,1,2,3$ correspond to singlet, triplet, quintet, and septet pairs.
We list possible indices in Tab.~\ref{tab:list}.

The time-dependent multipole functions are represented in terms of bogolon's physical quantities.
In order to see the contributions from the odd-frequency pairing of bogolons near the Fermi level, we define the odd-frequency multipoles $\tilde M^\eta$, $\tilde P^\eta$, $\tilde P^{\dg \eta}$ which are induced solely by the degrees of freedom near the Bogoliubov Fermi surfaces and are written in the forms
\begin{align}
  \tilde M^\eta_{\bm k}(\tau) &= C^\eta_{M}(\bm k) F_{\bm k}(\tau),
  \\
  \tilde P^\eta_{\bm k}(\tau) &= C^\eta_{P}(\bm k) F_{\bm k}(\tau),
  \\
  \tilde P^{\dg\eta}_{\bm k}(\tau) &= C^\eta_{P^\dg}(\bm k) F_{\bm k}(\tau).
\end{align}
The quantity $C$ is regarded as a kind of `susceptibility', showing how much of odd-frequency multipoles and multiplet pair of electrons are induced from the odd-frequency pair amplitude of bogolons.
We then define the quantities depending only on the rank
\begin{align}
  C^r_{X}(\bm k) &= \sqrt{
  \displaystyle
  \sum_{\eta \in r} |C_X^\eta(\bm k)|^2
  }
\end{align}
for $X=M,P,P^\dg$.
The result is shown in Fig.~\ref{fig:component} of the main text with a specific choice of the parameters.

\subsection*{2. Impurity potential}

The impurity scattering term is given in the language of the original electrons by
\begin{align}
  \mathscr H_{\rm imp} &= \sum_{i} \int \diff \bm r \sum_{\eta}
  U_\eta (\bm r - \bm R_i)
   \vec c^\dg (\bm r) \hat O^{\eta} \vec c (\bm r),
  \\
  \vec c(\bm r) &= \frac{1}{\sqrt{V}} \sum_{\bm k} \vec c_{\bm k} \, \epn^{\imu \bm k \cdot \bm r},
\end{align}
where we consider the isotropic ($\eta = 1$) and anisotropic ($\eta =xy,yz,zx,z^2,x^2-y^2$) scattering centers located at $\bm R_i$, both of which are electric degrees of freedom compatible with nonmagnetic impurities.
The full list of $4\times 4$ matrices are defined by using the $\hat {\bm J}$ matrix in Ref. \cite{Tamura20}.
The impurity position $\bm R_i$ is to be averaged.

The impurity potential can be rewritten in terms of bogolon's operators by using the relation
\begin{align}
  c_{\bm km} &= u_{\bm km}^* \al_{\bm k} + v_{-\bm k,m} \al_{-\bm k}^\dg,
\end{align}
where only the fermions that have Fermi surfaces are considered in the right-hand side.
We then obtain the impurity potential term for bogolons as
\begin{align}
  \mathscr H_{\rm imp} &= \frac{1}{V} \sum_{\bm{k} , \bm{q}} \rho_{\bm{q}} u_1(\bm{k} , \bm{q}) \alpha_{\bm{k} + \bm{q}}^\dagger \alpha_{\bm{k}} \notag \\
  &\quad + \frac{1}{V} \sum_{\bm{k} , \bm{q}} \rho_{\bm{q}} u_2(\bm{k} , \bm{q}) \alpha_{\bm{k} + \bm{q}}^\dagger \alpha_{-\bm{k}}^\dagger + \mathrm{H.c.} \notag \\
  &\quad + \mathrm{Const.},
\end{align}
where
\begin{align}
  u_1(\bm{k} , \bm{q}) &= \sum_{\eta} \sum_{m,m'} U_{\eta}(\bm{q}) [u_{\bm{k} + \bm{q} , m} O^\eta_{mm'} u_{\bm{k} , m'}^\ast \notag \\
  &\qquad - v_{\bm{k} , m}^\ast O_{mm'}^\eta v_{\bm{k} + \bm{q} , m'}], \\
  u_2(\bm{k} , \bm{q}) &= \sum_{\eta} \sum_{m,m'} U_{\eta}(\bm{q}) u_{\bm{k} + \bm{q} , m} O^\eta_{mm'} v_{-\bm{k} , m'}.
\end{align}

\subsection*{3. Single-particle spectral function}

We connect the single particle spectrum of the electrons to that of the bogolons.
We define the single-particle spectrum for $j=3/2$ electrons by
\begin{align}
  A^{\rm elec}_{\bm k}(\varepsilon) &= \frac{1}{2\pi f(\varepsilon)} \int_{-\infty}^{\infty} \diff t
  \la
  \vec \psi_{\bm k}^{\dg}
  \vec \psi_{\bm k} (t)
  \ra \epn^{\imu\varepsilon t},
\end{align}
where we have introduced the Nambu spinor $\vec\psi_{\bm k} = (\vec c_{\bm k}^{\rm T}, \vec c_{-\bm k}^{\dg})^{\rm T}$.
This is written in terms of bogolons, and near the Fermi level it has the form
\begin{align}
  A^{\rm elec}_{\bm k}(\varepsilon) &=
  \frac{1}{2\pi f(\varepsilon)} \int_{-\infty}^{\infty} \diff t
  \big[
  \la
  \al_{\bm k}^\dg
  \al_{\bm k} (t)
  \ra
  + \la
  \al_{-\bm k}
  \al_{-\bm k}^\dg (t) \ra
  \big] \epn^{\imu\varepsilon t}
  \notag \\
  &=
  - \frac{1}{\pi} \imag \trace \hat G_{\bm k}^{\rm R}(\varepsilon),
\end{align}
where the $2\times 2$ matrix $\hat G_{\bm{k}}^{\rm R}(\varepsilon)$ is the retarded version of the Green function defined in Eq.~\eqref{eq:G}.

\vspace{10mm}
\noindent
{\bf \large References}
\\[1mm]
See the list of references in the main text.

\end{document}